\documentclass[letterpaper,titlepage,11pt]{article}

\usepackage[T1]{fontenc}
\usepackage[utf8]{inputenc}
\usepackage{lmodern}
\usepackage{enumerate}
\usepackage[english]{babel}
\usepackage{comment}
\usepackage[numbers,sort&compress]{natbib}
\usepackage{bm}
\usepackage[usenames,dvipsnames]{xcolor}
\usepackage[a4paper]{geometry}
\usepackage[normalem]{ulem}

\usepackage[colorlinks=true,urlcolor=blue,citecolor=magenta]{hyperref}
\usepackage{amsmath,bbm}
\usepackage{amsfonts}
\usepackage{amssymb}
\usepackage{graphicx}
\usepackage[dvipsnames]{xcolor}
\usepackage{mathtools}
\usepackage{simplewick}
\usepackage{hyperref}
\usepackage{multirow}
\usepackage{enumitem}
\usepackage{mathrsfs}
\usepackage{physics}
\usepackage[compat=1.1.0]{tikz-feynman}
\usepackage{subcaption}
\usepackage{booktabs}
\usepackage{xcolor}

\usepackage{lipsum}

\usepackage{amsthm}

\allowdisplaybreaks[1]

\newcommand{\CJ}{\mathcal{J}}
\newcommand{\CM}{\mathcal{M}}

\newcommand{\CN}{\mathcal{N}}

\def\mN{\mathcal{N}}

\newcommand\blfootnote[1]{%
  \begingroup
  \renewcommand\thefootnote{}\footnote{#1}%
  \addtocounter{footnote}{-1}%
  \endgroup
}

\newcommand{\SU}{\text{SU}}

\newcommand{\spa}{\ , \ \ }
\newcommand{\nn}{\nonumber}
\newcommand{\psd}{\psi^\dagger}

\def\mba{\mathbf{a}}
\def\mbb{\mathbf{b}}
\def\mbc{\mathbf{c}}
\usepackage{etoolbox}

\def\le{\left(}
\def\ri{\right)}
\newcommand{\beq}{\begin{equation}}
\newcommand{\eeq}{\end{equation}}
\newcommand{\bea}{\begin{eqnarray}}
\newcommand{\eea}{\end{eqnarray}}

\def\ri{\right)}

\hypersetup{   colorlinks=true,  citecolor=blue, linkcolor=blue, urlcolor=red }

\begin{document}

\numberwithin{equation}{section}

\begin{titlepage}
\rightline{\vbox{   \phantom{ghost} }}

 \vskip 1.8 cm
\begin{center}
{\LARGE \bf
The Panorama of Spin Matrix Theory}
\end{center}
\vskip 1 cm

\title{}
\date{\today}
\author{Stefano Baiguera}
\author{Troels Harmark}
\author{Yang Lei}

\centerline{\large {{\bf Stefano Baiguera$^1$, Troels Harmark$^2$, Yang Lei$^{3}$}}}

\vskip 1.0cm

\begin{center}
\sl ${}^1$Department of Physics, Ben-Gurion University of the Negev, \\ David Ben Gurion Boulevard 1, Beer Sheva 84105, Israel \\[1mm]
\sl ${}^2$Niels Bohr International Academy, Niels Bohr Institute, University of Copenhagen,\\
Blegdamsvej 17, DK-2100 Copenhagen \O, Denmark\\[1mm]
\sl ${}^3$Institute for Advanced Study \& School of Physical Science and Technology,  \\ Soochow University, Suzhou 215006, P.R.~China
\end{center}

\vskip 1.3cm \centerline{\bf Abstract} \vskip 0.2cm 
\noindent
Spin Matrix theory describes near-BPS limits of $\mathcal{N}=4$ SYM theory, which enables us to probe finite $N$ effects like D-branes and black hole physics.
In previous works, we have developed the spherical reduction and spin chain methods to construct  Spin Matrix theory for various limits.
In this paper, by considering a supercharge $\mathcal{Q}$ which is cubic in terms of the letters, we construct the Hamiltonian of the largest Spin Matrix theory of $\mathcal{N}=4$ SYM, called the PSU$(1,2|3)$ Spin Matrix theory, as $H = \{\mathcal{Q}, \mathcal{Q}^\dagger \}$. 
We show the resulting Hamiltonian is automatically positive definite and manifestly invariant under supersymmetry.
The Hamiltonian is made of basic blocks which transform as supermultiplets. 
A novel feature of this Hamiltonian is its division into D-terms and F-terms that are separately invariant under PSU$(1,2|3)$ symmetry and positive definite. 
As all the other Spin Matrix theories arising from $\CN=4$ SYM can be acquired by turning off certain letters in the theory, we consider our work as revealing the ``Panorama'' of Spin Matrix theory.

\blfootnote{ \scriptsize{ \texttt{baiguera@post.bgu.ac.il, harmark@nbi.ku.dk, leiyang@suda.edu.cn}} }

\end{titlepage}
\newpage
\tableofcontents

%%%%%%%%%%%%%%%%%%%%%%%%%%%%%%%%%%%%
\section{Introduction}
\label{sec:introduction}

The AdS/CFT correspondence \cite{Maldacena:1997re}, which is a duality between four-dimensional $\mathcal{N}=4$ super Yang-Mills (SYM) theory with gauge group $\SU(N)$ and type IIB string theory on the background $\mathrm{AdS}_5 \times S^5$, is the best understood example of a concrete realization of the holographic principle, promising an understanding of how space, time and gravity emerge from a more fundamental quantum theory. 
However, observing this emergence is difficult as it requires a non-perturbative understanding of $\mathcal{N}=4$ SYM, especially for quantities that are not protected by supersymmetry. Exceptions include the planar limit $N\rightarrow \infty$ for which a powerful integrability symmetry appears \cite{Beisert:2010jr}. 
For capturing finite-$N$ effects, being non-perturbative in $1/N$, progress has in large part focussed on quantities protected by supersymmetry, for instance giant gravitons \cite{Biswas:2006tj} and the supersymmetric black hole \cite{Hosseini:2017mds,Cabo-Bizet:2018ehj,Choi:2018hmj,Benini:2018ywd} using index techniques.

In \cite{Harmark:2014mpa} a different approach to the non-perturbative, finite-$N$, regime of $\CN=4$ SYM is proposed: the Spin Matrix theory (SMT) limit of the AdS/CFT correspondence. This can be seen as a non-relativistic limit, 
both on the gauge and the string sides of the correspondence \cite{Harmark:2008gm,Harmark:2014mpa,Harmark:2017rpg,Harmark:2018cdl,Harmark:2020vll,Harmark:2019zkn,Baiguera:2020jgy,Baiguera:2020mgk,Baiguera:2021hky,Oling:2022fft}. In turn, it can also be seen as a near-BPS limit \cite{Harmark:2007px,Harmark:2014mpa,Harmark:2019zkn,Baiguera:2020jgy,Baiguera:2020mgk,Baiguera:2021hky,Oling:2022fft} and a regime that approaches a zero-temperature critical point \cite{Harmark:2014mpa,Harmark:2006di,Harmark:2006ta}. 
From such a limit arises Spin Matrix theory, which is a quantum-mechanical model that  selects a particular corner of the AdS/CFT correspondence for which the duality is more tractable \cite{Harmark:2014mpa}.
Indeed, a common reasoning in theoretical physics is that the study of limits of a certain system, such that some simplifications occur and analytic results are achievable, can be useful to understand the fundamental structure of a model and can be used a posteriori to learn information about the general case.

The Spin Matrix theories arise from SMT limits of  $\CN=4$ SYM, though they can also be defined in their own right without any reference to a parent theory. One way to view an SMT limit is that it approaches a BPS bound of $\CN=4$ SYM. 
More precisely, in the latter context one defines $\mathcal{N}=4$ SYM in the state-picture on $\mathbb{R} \times S^3$ and considers BPS bounds of the form
\beq
E \geq J \, , \qquad 
J \equiv a_1 \mathbf{S}_1 + a_2 \mathbf{S}_2 + b_1 \mathbf{Q}_1 + b_2 \mathbf{Q}_2 + b_3 \mathbf{Q}_3 \, ,
\label{eq:BPS_bounds}
\eeq
where $E$ is the energy, $a_i$ and $b_i$ are constant chemical potentials, $\mathbf{S}_i$ are the Cartan generators of rotations and $\mathbf{Q}_i$ the Cartan generators of $\mathrm{SU}(4)$ R-symmetry.
The restriction to a SMT is obtained by performing the decoupling limit
\beq
\lambda \rightarrow 0 \, , \qquad
\frac{1}{\lambda} \left( E - J  \right) \,\,\, \mathrm{finite}  \, , \qquad
N \,\,\, \mathrm{finite} \, ,
\label{eq:SMT_limits}
\eeq
where $\lambda=g^2 N$ is the 't Hooft coupling.
We stress that the general idea behind these decoupling limits is that they reduce the full $\mathcal{N}=4$ SYM theory to subsectors, where only tree-level and one-loop orders of the dilatation operator contribute in the partition function \cite{Harmark:2007px}.
Furthermore, only some of the modes of the original theory remain dynamical, while the others become infinitely heavy and effectively decouple from the Hamiltonian describing the near-BPS interactions.

In the present work we will consider the decoupling limit \eqref{eq:SMT_limits} with $a_1=a_2=b_1=b_2=b_3=1,$ which reduces $\mathcal{N}=4$ SYM to an SMT with PSU(1,2|3) symmetry.
The motivations to study PSU(1,2|3) SMT are several. 
Firstly, PSU(1,2|3) symmetry is the largest possible spin group that can arise from near-BPS limits of the form \eqref{eq:SMT_limits} and comprises as subsectors all the other allowed SMTs from $\CN=4$ SYM, as classified in \cite{Harmark:2007px}. The construction and symmetries of the effective Hamiltonian of various SMTs were studied in \cite{Harmark:2019zkn,Baiguera:2020jgy,Baiguera:2020mgk,Baiguera:2021hky}. 
%The PSU(1,2|3) SMT includes all these as subsectors, and realize the largest possible symmetry. 

Secondly, in \cite{Harmark:2016cjq} non-protected finite-$N$ effects of D-branes where matched using a strong-coupling limit of SU(2) SMT. This demonstrates SMT as a tool to obtain finite-$N$ effects. To approach black holes, one needs to use the PSU(1,2|3) SMT, as this is the only subsector that can capture BPS and near-BPS information about the SUSY macroscopic black hole in AdS$_5 \times S^5$ and its non-extremal generalizations \cite{Kunduri:2006ek,Chong:2005da,Chong:2005hr}.
That one can study finite-$N$ physics is due to $N$ being fixed in the above limit \eqref{eq:SMT_limits}.% 
\footnote{One can think of SMT as a generalization of a spin chain theory since SMT for $N = \infty$ reduces to a spin chain theory. Furthermore, perturbative $1/N$ corrections to $N = \infty$ can be interpreted as describing the dynamics of joining and splitting of spin chains in a gas of spin chains of various lengths \cite{Casteill:2007td}.
Other aspects of perturbative and non-perturbative effects in $1/N$, including the generalization of integrability, were further explored in the literature, see for example  \cite{Kristjansen:2010kg,Carlson:2011hy,deMelloKoch:2011dhs,Bargheer:2017nne}.}
Note that several works \cite{Chang:2022mjp,Grant:2008sk,Berkooz:2014uwa} have explored how to obtain the BPS limit of the SUSY black hole in AdS$_5 \times S^5$ by studying the one-loop dilatation operator of $\CN=4$ SYM, which is in direct correspondence with the Hamiltonian of PSU(1,2|3) SMT. Here we propose that in addition, one should also be able to capture near-BPS information about the near-extremal black hole.

On general grounds, any SMT limit of the form \eqref{eq:SMT_limits} maps the relativistic quantum field theory (QFT) $\mathcal{N}=4$ SYM into a quantum mechanical theory with non-relativistic traits, such as the emergence of a global $\mathrm{U}(1)$ symmetry interpreted as the conservation of mass or particle number. 
As a consequence, after the limit the anti-particles of the original QFT have decoupled.\footnote{In contrast, a relativistic QFT defined on $\mathbb{R} \times S^3$ admits anti-particles and therefore its particle number is not conserved.}
In this context, it would be highly interesting if one could find a field theory realization for the PSU(1,2|3) sector, similarly to cases that includes SU(1,1) symmetry in \cite{Harmark:2019zkn,Baiguera:2020jgy}.
The effective Hamiltonian describing their interactions in the near-BPS limit presents summation over positive modes only, instead of a full Fourier-like expansion.
Moreover, as already mentioned, their holographic duals are string theories with target space characterized by a Newton-Cartan geometry, and Galilean conformal algebra on the worldsheet.
Recent developments on this field can be found in \cite{Harmark:2017rpg,Bergshoeff:2018vfn,Bergshoeff:2018yvt,Gallegos:2019icg,Roychowdhury:2019sfo,Blair:2019qwi,Bergshoeff:2019pij,Yan:2019xsf,Gomis:2019zyu,Harmark:2019upf,Roychowdhury:2020yun,Kluson:2019xuo,Gomis:2020izd,Fontanella:2021hcb,Bidussi:2021ujm,Yan:2021lbe,Hartong:2021ekg,Oling:2022fft,Hartong:2022dsx,Harmark:2020vll,Harmark:2018cdl}.

\subsubsection*{Interacting Hamiltonian of Spin Matrix Theories}

SMTs can be defined starting from a Hilbert space with ladder operators transforming in the representation $R_s$ of a semi-simple Lie (super)-group $G_s$ and the adjoint representation of $\SU(N)$.
The Hilbert space is composed by all the possible harmonic oscillator states, created from the vacuum, which are singlets under the $R_s$ representation.
The interactions are described by a quartic Hamiltonian, built with two creation and two annihilation operators, invariant under 
all the generators of the spin group $G_s.$

There exists several techniques to compute the effective Hamiltonian in a certain near-BPS regime identified by the limit \eqref{eq:SMT_limits}:
\begin{enumerate}
\item One computes the loop corrections to the dilatation operator of $\mathcal{N} = 4$ SYM, and then zoom in towards the unitarity bound of interest. This method was extensively applied in \cite{Minahan:2002ve,Beisert:2002ff,Beisert:2003tq,Beisert:2003ys,Beisert:2004ry,Bellucci:2005vq,Bellucci:2006bv,Beisert:2007sk, Zwiebel:2007cpa,Beisert:2008qy,Beisert:2010jr}. 
One can extract an effective Hamiltonian in SMT language\footnote{By an effective Hamiltonian in SMT language, we mean that it is a quartic expression containing two creation and two annihilation operators, built using the fields which survive in the near-BPS limit \eqref{eq:SMT_limits}. } by looking at the action of the dilatation operator on spin chains and translating the results between different representations.
We will not pursue this approach in the present paper, but we will discuss its relation with the other methods below.
\item One performs an expansion in Kaluza-Klein modes along the three-sphere of $\mathcal{N}=4$ SYM defined on $\mathbb{R} \times S^3$.
This gives a classical Hamiltonian which is then directly promoted to a quantum-mechanical expression by requiring that no change of orderings is needed.
This approach was considered in \cite{Harmark:2019zkn,Baiguera:2020jgy,Baiguera:2020mgk,Baiguera:2021hky} and will be applied in Section \ref{sec:ham_sphere} for the PSU(1,2|3) sector.
We will refer to this procedure as {\sl{spherical expansion}}.
\item When the near-BPS limit preserves part of the original supersymmetry of the full symmetry group PSU(2,2|4), it is possible to define a cubic supercharge $\mathcal{Q}$ whose anticommutator closes into the interacting Hamiltonian of the sector:
\beq
\lbrace \mathcal{Q} , \mathcal{Q}^{\dagger} \rbrace_D = H_{\rm int} \, ,
\label{eq:intro_cubic_supercharge}
\eeq
where the subscript $D$ denotes the Dirac bracket.
This method is based on the observation that in the PSU(1,1|2) subsector there exists an enhanced psu(1,1)$^2$ subalgebra that can be used to represent the fermionic generators \cite{Beisert:2007sk, Zwiebel:2007cpa}, and was applied in the context of SMT in \cite{Baiguera:2021hky}.
We will use this technique as the starting point to derive the effective Hamiltonian of the PSU(1,2|3) sector in Section \ref{sec:ham_cubic_supercharge}.
\item One can build all the possible blocks (quadratic in the fields) that comprise an irreducible representation of the spin group $G_s$ characterizing the near-BPS limit \eqref{eq:SMT_limits}.
We then build the most general Hamiltonian quartic in the fields by combining the blocks determined in this way.
We applied successfully this technique in \cite{Baiguera:2020mgk}; in the case of PSU(1,2|3) sector, we will discuss this symmetry structure in Section \ref{sec:symmetry_structure}.  
\end{enumerate}

\begin{figure}[ht]
\centering
\includegraphics[scale=0.5]{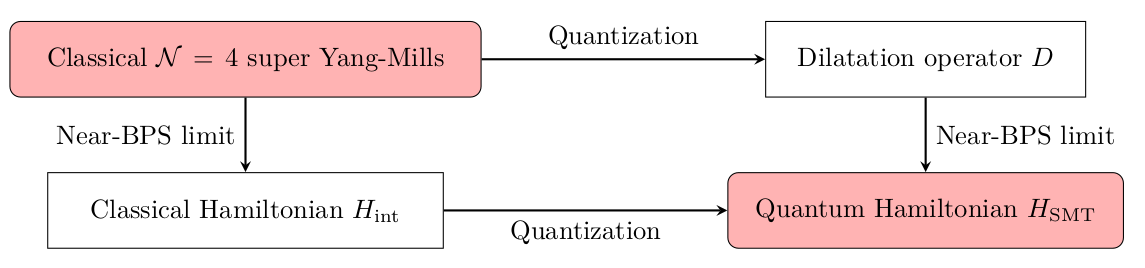}
\caption{\small Commutative diagram describing the relation between the methods 1 and 2 to go from the starting point, classical $\mathcal{N}=4$ SYM (top left), to the final result, an effective quantum SMT Hamiltonian (bottom down).
According to method 1, we move right-down by computing one-loop corrections to the dilatation operator and then restricting to a near-BPS limit \eqref{eq:SMT_limits}.
Following method 2, we move down-right by performing the sphere expansion procedure and then giving a recipe to quantize the theory.}
\label{fig:commutative_diagram}
\end{figure}

We discuss the advantages and disadvantages of the previous methods.
First of all, it can be shown that they are all equivalent, in that the quantum interacting Hamiltonian obtained applying any of them is the same.
This is particularly non-trivial by observing that the techniques 1 and 2 consist in reversing the order of two limits: the near-BPS decoupling implemented using the prescription \eqref{eq:SMT_limits}, and the quantization procedure (which can be formally interpreted as performing the limit $\hbar \rightarrow 0$).
In this regard, we have shown for several spin groups $G_s$ that these procedures commute, {\sl i.e.}, the diagram in fig.~\ref{fig:commutative_diagram} is commutative \cite{Harmark:2019zkn,Baiguera:2020jgy,Baiguera:2020mgk,Baiguera:2021hky}.
While we will not show it explicitly, these results provide non-trivial expectations that this phenomenon will happen in the present PSU(1,2|3) sector, too.

\noindent
The main advantage of the procedures 1 and 2 is that they are conceptually straightforward, since they provide a way to extract the Hamiltonian for any given sector by applying a precise algorithm.
The disadvantage is that they are technically complicated, and therefore it becomes really involved to perform the explicit computations when the spin group $G_s$ is big.
The strategy 4 is elegant because it identifies a class of fundamental blocks that are irreducible representations of the algebra and can be used to build all the interactions. 
On the other hand, in sectors containing F-terms in the Hamiltonian, the identification of the fundamental blocks becomes rather difficult, as was observed for the PSU(1,1|2) subsector \cite{Baiguera:2021hky}.
Since the PSU(1,2|3) sector will also include this kind of interactions, we face the same difficulty here.

Nonetheless, we will be able to find the symmetry structure with the help of the method 3.
Its disadvantage is that the procedure cannot be applied in the absence of supersymmetry and it also fails in all the sectors without the singlet fermionic letter denoted as\footnote{See Section \ref{sec:preliminaries} for the notation of the letters in the PSU(1,2|3) theory.} $|\chi_{n,k} \rangle .$ 
However, since this letter and its descendants are part of the field content of the PSU(1,2|3) sector, we will be able to perform the computation.
This method is simpler than the others because it only requires to identify a cubic fermionic generator invariant under all the residual symmetries in the near-BPS limit, instead of working directly at the level of the quartic Hamiltonian.
Once the cubic supercharge $\mathcal{Q}$ is fixed, the recipe 
 \eqref{eq:intro_cubic_supercharge} provides a straightforward way to obtain the interactions.
Another advantage is that supersymmetry invariance guarantees that the Hamiltonian computed in this way is positive-definite.

The paper is organized as follows.
We introduce the field content of the theory and the generators of the PSU(1,2|3) algebra in Section \ref{sec:preliminaries}. 
The main technique that we will adopt in the present work is based on the cubic supercharge method, that we will present in Section \ref{sec:ham_cubic_supercharge}.
The interacting Hamiltonian naturally organizes into a positive-definite expression built out of quadratic blocks in the fields. This allows for a symmetry analysis of the result in Section \ref{sec:symmetry_structure}.  
In Section \ref{sec:ham_sphere}, we use the spherical expansion procedure as a consistency check and as an input to uniquely fix the interacting Hamiltonian.
We conclude in Section \ref{sec:discussion}.
Appendices are reserved for additional technical details: the invariance of the cubic supercharge under the PSU(1,2|3) generators in Appendix \ref{app:details_invariance_cubic}, and the notation for the spherical expansion in Appendix \ref{app:sphere_red}.

\section{Preliminaries}
\label{sec:preliminaries}

In this Section, we will introduce the letters of PSU$(1,2|3)$ Spin Matrix theory. 
Our conventions follow \cite{Harmark:2007px}.

\begin{figure}
	\centering
	\includegraphics[trim=3cm 12cm 10cm 4cm,width=0.35\linewidth]{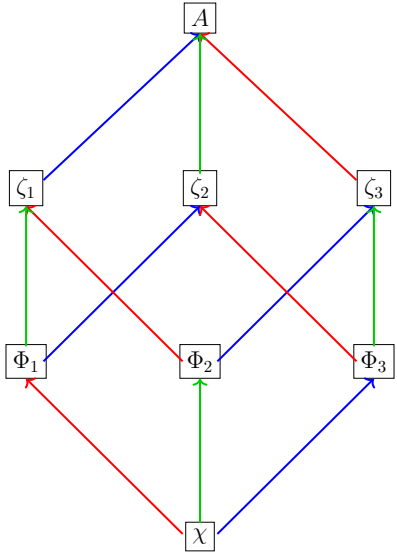}
	\caption{The actions by the supercharges $S_a^\dagger$ in eq.~\eqref{eq:oscillator_supercharges} are shown explicitly by colored arrows. The blue, green and red arrows refer to the generators $S_{1,2,3}^\dagger$, respectively. The $S_a$ operations are represented by inverse arrows. We can think of the diagram as a cube with 8 nodes and 12 edges. Each edge represents an $\mN=1$ supermultiplet while each surface represents an $\mN=2$ supermultiplet.}
	\label{fig:letter123s}
\end{figure}

The $\mN=4$ SYM theory contains six scalars $\Phi_a$, 16 complex  fermions $\chi$ and $\bar{\chi}$, one gauge field $A$ and four derivative letters $d_i\,,\bar{d}_i$ with $i=1,2$. 
The decoupling condition for the fields of the PSU$(1,2|3)$ subsector of $\mN=4$ SYM follows directly from the BPS condition: 
\begin{equation}
	E_0= \mathbf{Q}_1+\mathbf{Q}_2+\mathbf{Q}_3+ \mathbf{S}_1+\mathbf{S}_2 \, .
\end{equation}
where $E_0$ is the bare energy. 
There are five fermions $\chi_{1,2},\bar{\chi}_{3,5,7}$, three scalars $\Phi_{1,2,3}$, one gauge component of $A$ and two derivative letters $d_{1,2}$ satisfying this condition. 
Among these letters, the chiral fermions $\chi_{1,2}$ are subject to the Dirac equation \cite{Kinney:2005ej}
\begin{equation}
	d_1\chi_2 -d_2\chi_1 =0\,.
\end{equation}
We can then introduce an {\sl{ancestor}} fermion $\chi$ such that $\chi_{1,2}$ are the descendants of the $\chi$ field, defined as 
\begin{equation}
	\chi_{i} =d_{i} \chi\,,\qquad i=1,2 \, .
	\label{eq:ancestor_fermion}
\end{equation}
For notational convenience, we denote the anti-chiral fermions $\bar{\chi}_{3,5,7}$ as 
\begin{equation}
	\zeta_1= \bar{\chi}_3, \qquad \zeta_2 = \bar{\chi}_5,\qquad \zeta_3 =\bar{\chi}_7 \, .
\end{equation}
The oscillator representation for the $\mathrm{u}(2,2|4)$ algebra of $\mathcal{N}=4$ SYM is built by introducing two sets of bosonic oscillators $\mathbf{a}_{\alpha}, \mathbb{b}_{\dot{\alpha}}$ with $\alpha, \dot{\alpha} \in \lbrace 1,2\rbrace$ and one set of fermionic operators $\mathbf{c}_a$ with $a \in \lbrace 1,2,3,4 \rbrace$ whose commutation relations read
\beq
[\mathbf{a}_{\alpha}, \mathbf{a}^{\dagger}_{\beta}] = \delta_{\alpha\beta} \, , \qquad
[\mathbf{b}_{\dot{\alpha}}, \mathbf{b}^{\dagger}_{\dot{\beta}}] = \delta_{\dot{\alpha} \dot{\beta}} \, , \qquad
\lbrace \mathbf{c}_{a}, \mathbf{c}^{\dagger}_b \rbrace = \delta_{ab} \, .
\eeq
The restriction to the PSU(1,2|3) group is achieved by setting $\mathbf{b}_{2} \mathbf{b}^{\dagger}_2 =0$ and $\mathbf{c}_{4} \mathbf{c}^{\dagger}_4 =1.$
The oscillator realization of the SU$(1,2)$  generators is given by \cite{bars1990unitary}
\begin{align} \label{eq:newbasisforSU21}
	\begin{split}
		L_0 &= \frac{1}{2} (1+\mba_1^\dagger \mba_1 +\mbb_1^\dagger \mbb_1), \quad L_+ = \mba_1^\dagger \mbb_1^\dagger, \quad L_{-} =\mba_1 \mbb_1 \, , \\ 
		\tilde{L}_0 &= \frac{1}{2} (1+\mba_2^\dagger \mba_2 +\mbb_1^\dagger \mbb_1), \quad \tilde{L}_+ = \mba_2^\dagger \mbb_1^\dagger, \quad \tilde{L}_{-} = \mba_2 \mbb_1 \, , \\ 
		J_+ &= \mba_1^\dagger \mba_2, \qquad J_- =\mba_2^\dagger \mba_1 \, .
	\end{split}
\end{align}
and the supercharges are 
\begin{equation}
	Q_a = \mba_1 \mbc_a^\dagger, \qquad \tilde{Q}_a = \mba_2 \mbc_a^\dagger, \qquad S_a = \mbb_1 \mbc_a, \qquad a=1,2,3 \, ,
	\label{eq:oscillator_supercharges}
\end{equation}
plus their hermitian conjugates.
Finally, the SU$(3)$ R-symmetry generators are
\begin{align}\label{eq:R-symmetry-generators}
	\begin{split}
&	R_0 = \frac{1}{2}(\mbc_3^\dagger \mbc_3-\mbc_2^\dagger \mbc_2) \spa R_+ = \mbc_3^\dagger \mbc_2 \spa R_- = \mbc_2^\dagger \mbc_3 \\
&	\tilde{R}_0 = \frac{1}{2}(\mbc_2^\dagger \mbc_2 - \mbc_1^\dagger \mbc_1) \spa \tilde{R}_+ = \mbc_2^\dagger \mbc_1 \spa \tilde{R}_- = \mbc_1^\dagger \mbc_2
\\
&	T_+ = \mbc_3^\dagger \mbc_1 \spa\qquad  T_- = \mbc_1^\dagger \mbc_3 \, .
\end{split}
\end{align}
Thus the letters satisfying the decoupling condition are
\begin{eqnarray}\label{eq:statesSU1232}
	\ket{\chi_{n,k}} &=& \frac{1}{\sqrt{n!k!(n+k-1)!}} (\mba_1^\dagger )^n   (\mba_2^\dagger )^k( \mbb_1^\dagger )^{n+k-1} \mbc_4^\dagger  \ket{0} \\
	\ket{\Phi^1_{n,k}} &=&  \frac{1}{\sqrt{n!k!(n+k)!}}  (\mba_1^\dagger\mbb_1^\dagger)^n (\mba_2^\dagger \mbb_1^\dagger)^k \mbc_3^\dagger \mbc_4^\dagger \ket{0} \\ 
	\ket{\Phi^2_{n,k}} &=&  \frac{1}{\sqrt{n!k!(n+k)!}}  (\mba_1^\dagger\mbb_1^\dagger)^n (\mba_2^\dagger \mbb_1^\dagger)^k \mbc_2^\dagger \mbc_4^\dagger \ket{0} \\ 
	\ket{\Phi^3_{n,k}} &=&  \frac{1}{\sqrt{n!k!(n+k)!}}  (\mba_1^\dagger\mbb_1^\dagger)^n (\mba_2^\dagger \mbb_1^\dagger)^k \mbc_1^\dagger \mbc_4^\dagger \ket{0} \\ 
	\ket{\zeta^1_{n,k}} &=& \frac{1}{\sqrt{n!k!(n+k+1)!}} (\mba_1^\dagger \mbb_1^\dagger)^n   (\mba_2^\dagger \mbb_1^\dagger)^k \mbb_1^\dagger  \mbc_1^\dagger  \mbc_2^\dagger  \mbc_4^\dagger  \ket{0}\\
	\ket{\zeta^2_{n,k}} &=& \frac{1}{\sqrt{n!k!(n+k+1)!}} (\mba_1^\dagger \mbb_1^\dagger)^n   (\mba_2^\dagger \mbb_1^\dagger)^k \mbb_1^\dagger  \mbc_1^\dagger  \mbc_3^\dagger  \mbc_4^\dagger  \ket{0}  \\ 
	\ket{\zeta^3_{n,k}} &=& \frac{1}{\sqrt{n!k!(n+k+1)!}} (\mba_1^\dagger \mbb_1^\dagger)^n   (\mba_2^\dagger \mbb_1^\dagger)^k \mbb_1^\dagger  \mbc_2^\dagger  \mbc_3^\dagger  \mbc_4^\dagger  \ket{0} \\
	\ket{A_{n,k}} &=&  \frac{1}{\sqrt{n!k!(n+k+2)!}} (\mba_1^\dagger \mbb_1^\dagger)^n   (\mba_2^\dagger \mbb_1^\dagger)^k (\mbb_1^\dagger)^2 \mbc_1^\dagger  \mbc_2^\dagger  \mbc_3^\dagger  \mbc_4^\dagger  \ket{0} \, .
\end{eqnarray}
As an example of the actions of the supercharges, we have shown the action of $S_a$ and $S_a^\dagger$ in figure \ref{fig:letter123s}.
We will use $V_I$ to label the letters, where $I=0,1,2,3$ correspond to $\{\chi,\Phi_a,\zeta_a,A \}$, respectively. 
We can write the
SU$(1,2)$ symmetry generators in terms of letters of the PSU$(1,2|3)$ sector as
\begin{equation}\label{eq:SU12-generator-bysusy}
	\begin{array}{rl}
		L_+ &= \sum_{I=1}^3 \sum_{n,k=0}^\infty \sqrt{(n+1)(n+k+I)} \tr \Big[ \big( V_I^\dagger\big)_{n+1,k} \big( V_I\big)_{n,k} \Big] \, , \\[2mm]
		L_0 &= \sum_{I=1}^3 \sum_{n,k=0}^\infty \left(n+\frac{k+I}{2} \right)\tr\Big[ \big( V_I^\dagger\big)_{n,k} \big( V_I\big)_{n,k} \Big] \, , \\[2mm]
		L_- &= \sum_{I=1}^3 \sum_{n,k=0}^\infty \sqrt{n(n+k+I-1)} \tr \Big[\big( V_I^\dagger\big)_{n-1,k} \big( V_I\big)_{n,k} \Big] \, ,  \\[2mm]
		\tilde{L}_+ &= \sum_{I=1}^3 \sum_{n,k=0}^\infty \sqrt{(k+1)(n+k+I)}\tr \Big[ \big( V_I^\dagger\big)_{n,k+1} \big( V_I\big)_{n,k} \Big] \, , \\[2mm]
		\tilde{L}_0 &= \sum_{I=1}^3 \sum_{n,k=0}^\infty \left(k+\frac{n+I}{2} \right) \tr \Big[\big( V_I^\dagger\big)_{n,k} \big( V_I\big)_{n,k} \Big] \, , \\[2mm]
		\tilde{L}_- &= \sum_{I=1}^3 \sum_{n,k=0}^\infty \sqrt{k(n+k+I-1)} \tr \Big[\big( V_I^\dagger\big)_{n,k-1} \big( V_I\big)_{n,k} \Big] \, , \\[2mm]
		J_+ &= \sum_{I=1}^3 \sum_{n,k=0}^\infty \sqrt{k(n+1)} \tr \Big[\big( V_I^\dagger\big)_{n+1,k-1} \big( V_I\big)_{n,k}  \Big] \, , \\[2mm]
		J_- &= \sum_{I=1}^3 \sum_{n,k=0}^\infty \sqrt{n(k+1)} \tr \Big[\big( V_I^\dagger\big)_{n-1,k+1} \big( V_I\big)_{n,k} \Big] \, .
	\end{array}
\end{equation} 
The representation theory of SU$(1,2)$ was worked out in \cite{bars1990unitary} and reviewed in \cite{Baiguera:2020mgk}.
Here we only emphasize the crucial properties needed for this work:
\begin{itemize}
	\item While the representations of the SU$(1,1)$ algebra are labelled by a quantum number $j$ which parametrizes the quadratic Casimir as $C_2=-j(j+1)$, similarly the representations of the SU$(1,2)$ algebra are labelled by two quantum numbers $(p,q)$, which are related to the quadratic and cubic Casimirs as 
	\begin{align}\label{eq:twoCasimir-SU12}
		\begin{split}
		C_2 &=  p+q +\frac{1}{3} (p^2+pq+q^2) \, , \\
		C_3 &= \frac{1}{27} (p-q)(p+2q+3) (q+2p+3)  \, .
		\end{split}
	\end{align}
\item In this paper, the representation relevant to us is the \emph{integer} series with $(p,q)=(0,I-3)$ for $I=0,1,2,3$.
The algebra action \eqref{eq:SU12-generator-bysusy} on the letters transforming in these representations are 
\begin{equation}\label{eq:SU122-block-I}
	\begin{array}{rl}
		(L_+)_D(V_I^\dagger)_{n,k}  &=  \sqrt{(n+1)(n+k+I)} (V_I^\dagger)_{n+1,k}\,, \\[1mm]
		(\tilde{L}_+)_D(V_I^\dagger)_{n,k}  &=  \sqrt{(k+1)(n+k+I)} (V_I^\dagger)_{n,k+1}\,, \\[1mm]
		(L_0)_D(V_I^\dagger)_{n,k}  &=  \left(n+\frac{k+I}{2} \right) (V_I^\dagger)_{n,k}\,, \\[1mm] 	
		(\tilde{L}_0)_D(V_I^\dagger)_{n,k} &=  \left(k+\frac{n+I}{2} \right) (V_I^\dagger)_{n,k}\,, \\[1mm]
		(L_{-})_D(V_I^\dagger)_{n,k}  &=  \sqrt{n(n+k+I-1)} (V_I^\dagger)_{n-1,k}\,,  \\[1mm]
		(\tilde{L}_{-})_D(V_I^\dagger)_{n,k} &=  \sqrt{k(n+k+I-1)} (V_I^\dagger)_{n,k-1}\,, \\[1mm]
		(J_+)_D(V_I^\dagger)_{n,k}   &= \sqrt{k(n+1)} (V_I^\dagger)_{n+1,k-1}\,, \\[1mm]
		(J_-)_D(V_I^\dagger)_{n,k}   &= \sqrt{n(k+1)} (V_I^\dagger)_{n-1,k+1} \, .
	\end{array}
\end{equation}
\end{itemize}
We can therefore check that the letters $V_I$ transforms in the $(p,q)=(0,I-3)$ representations of SU$(1,2)$.

The remaining generators of the PSU(1,2|3) can also be written in the previous representation.
The supercharges read
\begin{align} \label{eq:supercharge-asletters}
	\begin{split}
Q_{4-a} &= \sum_{n=1}^\infty \sum_{k=0}^\infty  \sqrt{n}\, \tr \left( (\Phi_a^\dagger)_{n-1,k}\, \chi_{n,k} + \epsilon_{abc} (\zeta_b^\dagger)_{n-1,k} (\Phi_c)_{n,k} +  A_{n-1,k}^\dagger (\zeta_a)_{n,k} \right) \\
\tilde{Q}_{4-a} &=  \sum_{n=0}^\infty \sum_{k=1}^\infty  \sqrt{k}\, \tr \left( (\Phi_a^\dagger)_{n,k-1} \chi_{n,k} + \epsilon_{abc} (\zeta_b^\dagger)_{n,k-1} (\Phi_c)_{n,k} +  A_{n,k-1}^\dagger (\zeta_a)_{n,k} \right) \\ 
S_{4-a} &= \sum_{n,k=0}^\infty \tr \left( \sqrt{n+k} \, \chi^\dagger_{n,k} (\Phi_a)_{n,k} - \sqrt{n+k+1} \, \epsilon_{abc} (\Phi_b^\dagger)_{n,k} (\zeta_c)_{n,k}  \right. \\
&  \left. + \sqrt{n+k+2} \, (\zeta_a^\dagger)_{n,k} A_{n,k} \right) \, .
\end{split}
	\end{align}
The $\SU(3)$ R-symmetry generators in terms of letters are: 
\begin{align}\label{eq:R-symmetry-letters}
	\begin{split}
&	R_0 = \sum_{n,k=0}^\infty \tr  \left(  \frac{1}{2}  (\Phi_1^\dagger)_{n,k} (\Phi_1)_{n,k} - \frac{1}{2}  (\Phi_2^\dagger)_{n,k} (\Phi_2)_{n,k}  + \frac{1}{2} (\zeta_2^\dagger)_{n,k} (\zeta_2)_{n,k} -\frac{1}{2} (\zeta_1^\dagger)_{n,k} (\zeta_1)_{n,k}  \right) \\
& 	R_+ = \sum_{n,k=0}^\infty \tr  \left((\Phi_1^\dagger)_{n,k}  (\Phi_2)_{n,k}-(\zeta_2^\dagger)_{n,k} (\zeta_1)_{n,k} \right)
\spa
R_- = \sum_{n,k=0}^\infty \tr  \left( (\Phi_2^\dagger)_{n,k} (\Phi_1)_{n,k} - (\zeta_1^\dagger)_{n,k} (\zeta_2)_{n,k}\right) \\
& 	\tilde{R}_0 = \sum_{n,k=0}^\infty \tr  \left(  \frac{1}{2}  (\Phi_2^\dagger)_{n,k} (\Phi_2)_{n,k} - \frac{1}{2}  (\Phi_3^\dagger)_{n,k} (\Phi_3)_{n,k} -\frac{1}{2} (\zeta_2^\dagger)_{n,k} (\zeta_2)_{n,k}+ \frac{1}{2} (\zeta_3^\dagger)_{n,k} (\zeta_3)_{n,k}  \right) \\
& 	\tilde{R}_+ = \sum_{n,k=0}^\infty \tr  \left( (\Phi_2^\dagger)_{n,k} (\Phi_3)_{n,k} - (\zeta_3^\dagger)_{n,k} (\zeta_2)_{n,k} \right)
\spa
\tilde{R}_- = \sum_{n,k=0}^\infty \tr  \left( (\Phi_3^\dagger)_{n,k} (\Phi_2)_{n,k} - (\zeta_2^\dagger)_{n,k}  (\zeta_3)_{n,k}
\right) \\
& 	T_+ = \sum_{n,k=0}^\infty \tr  \left( (\Phi_1^\dagger)_{n,k} (\Phi_3)_{n,k}  - (\zeta_3^\dagger)_{n,k} (\zeta_1)_{n,k}\right)
\spa
T_- = \sum_{n,k=0}^\infty \tr  \left( (\Phi_3^\dagger)_{n,k} (\Phi_1)_{n,k}  -(\zeta_1^\dagger)_{n,k} (\zeta_3)_{n,k}
\right) \, .
	\end{split}
\end{align}
The fields are defined in such a way that their Dirac brackets have a standard normalization \cite{Baiguera:2020mgk}
\begin{align}
& \{ ( \chi_{n,k} )^i{}_j , (\chi^\dagger_{n',k'})^m{}_l \}_D = \{ ( \chi^\dagger_{n,k} )^i{}_j , (\chi_{n',k'})^m{}_l \}_D = \delta_{n,n'} \delta_{k,k'} \delta^i_l \delta^m_j \, , &
\label{eq:Dirac_brackets_singl_ferm} \\
& \{ ( (\Phi_a)_{n,k} )^i{}_j , ((\Phi_b^\dagger)_{n',k'})^m{}_l \}_D = - \{ ( (\Phi_a^\dagger)_{n,k} )^i{}_j , ((\Phi_b)_{n',k'})^m{}_l \}_D = \delta_{a,b} \delta_{n,n'} \delta_{k,k'} \delta^i_l \delta^m_j \, ,& \label{eq:Dirac_brackets_scalar}  \\
& \{ ( (\zeta_a)_{n,k} )^i{}_j , ((\zeta_b^\dagger)_{n',k'})^m{}_l \}_D = \{ ( (\zeta_a^\dagger)_{n,k} )^i{}_j , ((\zeta_b)_{n',k'})^m{}_l \}_D = \delta_{a,b} \delta_{n,n'} \delta_{k,k'} \delta^i_l \delta^m_j \, , &
\label{eq:Dirac_brackets_tripl_ferm} \\
& \{ ( A_{n,k} )^i{}_j , (A^\dagger_{n',k'})^m{}_l \}_D = -\{ ( A^\dagger_{n,k} )^i{}_j , (A_{n',k'})^m{}_l \}_D =  \delta_{n,n'} \delta_{k,k'} \delta^i_l \delta^m_j \, , &
\label{eq:Dirac_brackets_gauge} 
\end{align}
where we explicitly indicated the index structure under the $\SU(N)$ colour group, too.

%%%%%%%%%%%%%%%%%%%%%
\section{Hamiltonian from cubic supercharge}
\label{sec:ham_cubic_supercharge}

In this Section we compute the SMT Hamiltonian of the PSU(1,2|3) sector using the cubic supercharge method (labelled by 3 in Section \ref{sec:introduction}).
It consists of constructing the most generic cubic fermionic generator $\mathcal{Q}$ invariant under the full PSU(1,2|3) symmetry group, as we will do in Section \ref{ssec:construction_cubic_supercharge}.
Afterwards, the cubic supercharge is used to derive the interactions by using the identity \eqref{eq:intro_cubic_supercharge}.
One of the main features of the effective theories derived in the near-BPS limit is that they are expected to be positive definite, as a consequence of describing the reaction of a physical system in departing from the point in parameter space where the BPS bound \eqref{eq:BPS_bounds} is saturated. 
More specifically, one can interpret the effective Hamiltonian as a distance in the linear space of the representation identified by a set of fundamental blocks \cite{Baiguera:2020mgk}.
In this regard, the method applied here is particularly convenient because a Hamiltonian built using the identity \eqref{eq:intro_cubic_supercharge} is automatically positive-definite, as a consequence of supersymmetry invariance.
We will show in Section \ref{ssec:derivation_Hint_cubic} that it is possible to identify a block structure by inspection of the interacting Hamiltonian, thus providing a more direct way to show the positivity of the spectrum.

%%%%%%%%%%%%%%%%%%%
\subsection{Construction of the cubic supercharge}
\label{ssec:construction_cubic_supercharge}

In order to build a cubic fermionic generator $\mathcal{Q}$ invariant under the full PSU(1,2|3) spin group, we decompose it as a linear combination of terms 
\beq
\mathcal{Q} = \sum_{A} \alpha_A T_A \, ,
\label{eq:generic_linear_combination_seccubic}
\eeq
where $\alpha_A$ are real coefficients.
We require that
\begin{itemize}
\item Each term $T_A$ is a singlet under the adjoint representation of the colour group $\SU(N)$ and contains two raising and one lowering operators.\footnote{This is required in order to obtain an interacting Hamiltonian with two raising and two lowering operators via the application of eq.~\eqref{eq:intro_cubic_supercharge}.}
\item Each term $T_A$ is invariant under the bosonic spin subgroup $\SU(1,2) \times \SU(3)$, {\sl i.e.}, it commutes with all the bosonic generators.
This will classify all the structures allowed to enter the linear combination \eqref{eq:generic_linear_combination_seccubic}. 
\item The full linear combination \eqref{eq:generic_linear_combination_seccubic} is invariant under all the fermionic generators of the PSU(1,2|3) group.
This part of the procedure fixes the relative coefficients among the structures $T_A.$ 
\end{itemize}

We begin by addressing the correct structure with respect to the $\SU(N)$ adjoint representation. 
This requires to contract all the colour indices and to collect two of the fields into an (anti-)commutator structure.
We point out the general property\footnote{In the remaining part of the paper, we will denote the Lie parenthesis in two different ways. We will use $\lbrace \cdot , \cdot \rbrace_D$ for the Dirac brackets involving fields with any statistics, while we will use the notation $[\cdot , \cdot \rbrace $ without subscript to refer to the matrix (anti)commutators of the $\mathrm{SU}(N)$ colour group.   
}
\begin{equation}
\tr ( A [B,C\} ) = \tr ( [A,B\} C ) \, ,
\label{eq:identity_traces}
\end{equation}
where $A$, $B$ and $C$ can be either c-valued or Grassmann-valued fields in the adjoint representation of $\SU(N)$, and where $[\cdot,\cdot\}$ represents a commutator or an anti-commutator depending on the parity of the generators.
Due to the identity \eqref{eq:identity_traces}, it is not restrictive to only consider cubic combinations of fields of the form written in the right-hand side.

\subsubsection*{Invariance under the bosonic generators}

We begin by focusing on the generators of the $\SU(1,2)$ bosonic subgroup, which is responsible for assigning the integer numbers $(n,k)$  to any field in the theory.
Given three fields $V$, $\tilde{V}$ and $\hat{V},$ any cubic generator containing two creation operators and one annihilation operator can be a singlet in the adjoint representation of $\SU(N)$ and under the $\SU(1,2)$ spin subgroup only if it takes the form
\begin{equation}
\tr ( [V^\dagger_{n,k} , \tilde{V}^\dagger_{n',k'} \} \hat{V}_{n+n',k+k'} ) \, .
\end{equation}
Indeed, the invariance under the $L_0$ generator implies that the summation of labels $(n,k)$ involving the annihilation fields needs to match the summation of labels of the hermitian conjugate fields, which represent instead creation operators.
This statement simply corresponds in the language of spherical expansion (that will be analyzed in Section \ref{sec:ham_sphere}) to the conservation of momenta.

In the following steps, the index $I$ corresponding to the labelling of letters introduced in Section \ref{sec:preliminaries}, and we will make use of eq.~\eqref{eq:SU122-block-I}.
Assuming that the field $V$ transforms in the representation labelled by $I=i$  and that $\tilde{V}$ transforms instead with $I=j$, we further need to impose that $\hat{V}$ transforms with $I=i+j$ to obtain a cubic combination commuting with the generator $L_0.$
More precisely, this restricts the form of the generic $T_A$ term in eq.~\eqref{eq:generic_linear_combination_seccubic} to be
\begin{equation}
\label{eq:generic_TA_term}
T_A = \sum_{n,k,n',k'=0}^\infty P^{(i,j)}_{n,k,n',k'} \tr ( [  V^\dagger_{n,k}  \tilde{V}^\dagger_{n',k'} \} \hat{V}_{n+n',k+k'}) \, ,
\end{equation}
with
\begin{equation}
P^{(i,j)}_{n,k,n',k'} = \sqrt{\frac{(k+n+i-1)!(k'+n' + j-1)!(n+n')!(k+k')!}{(k+k'+n+n' + i+j-1)!n!k!n'!k'!}} \, .
\label{eq:Pij_coefficients}
\end{equation}
These coefficients account for the symmetry properties of the generators under the $\SU(1,2)$ subgroup; from the point of view of the spherical expansion procedure, they can be related to Clebsch-Gordan coefficients defined on the three-sphere:
\begin{align}
	\begin{split}
 C^{\frac{1}{2}(n+n'+k+k'),\frac{1}{2}(k+k'-n-n')}_{\frac{1}{2}(n'+k'),\frac{1}{2}(k'-n');\, \frac{1}{2}(n+k),\frac{1}{2}(k-n) } &= \sqrt{\frac{(n+k)!(n'+k')!(n+n')!(k+k')!}{(n+n'+k+k')!n!k!n'!k'!}} \\
&= \sqrt{n+k} P^{(0,1)}_{n,k;n',k'}
	\end{split}
	\label{eq:relation_P_CG}
\end{align} 
The PSU(1,2|3) invariance secretely encodes information about the geometric structure of the underlying theory.
At this point, one can check by direct computation that the object defined in eq.~\eqref{eq:generic_TA_term} is invariant under all the generators of the $\SU(1,2)$ subgroup.
Indeed, we obtain
\beq
\lbrace L_+, T_A \rbrace_D = 
\lbrace L_-, T_A \rbrace_D = 
\lbrace J_+, T_A \rbrace_D =  0 \, .
\eeq
Since the expression for $T_A$ is symmetric in the indices $(n,k),$ this easily implies that the following identities are also true:
\beq
\lbrace \tilde{L}_+, T_A \rbrace_D = 
\lbrace \tilde{L}_-, T_A \rbrace_D = 
\lbrace J_-, T_A \rbrace_D =  0 \, .
\eeq
Further details on the $\SU(1,2)$ invariance of $T_A$ are given in Appendix \ref{app:bos_generators}.

\vskip 5mm

\noindent
Given the generic ansatz \eqref{eq:generic_TA_term}, we move on to classify all the possible set of fields that can enter such expression.
Thus we further impose the invariance under the $\SU(3)$ R-symmetry.
In particular, this implies that the total eigenvalue of the Cartan generators should vanish, {\sl i.e.}, we require $R_0=\tilde{R}_0=0$.
We systematically approach the classification of terms in the following way: 
\begin{itemize}
\item First, we fix the value $i=0$ of the representation under which the field $V$ transforms.
\item Given the constraint $i+j \leq 3$ for the field $\hat{V},$ we consider all the possible values for the representation $j$ of $\tilde{V}.$
For each case, we construct the $\SU(3)$ invariant combination.
\item We repeat the procedure by increasing the integer $i$ by one unit, until the maximal value $i=3.$
\end{itemize}

\noindent
Working in this way, one can show that all the possible terms are given by
\begin{equation}
\label{T1}
T_1= \frac{1}{2} \sum_{n,k,n',k'=0}^\infty P^{(0,0)}_{n,k,n',k'} \tr ( \chi^\dagger_{n,k} \{ \chi^\dagger_{n',k'}, \chi_{n+n',k+k'} \} ) \, ,
\end{equation}
\begin{equation}
\label{T2}
T_2=\sum_{n,k,n',k'=0}^\infty P^{(0,1)}_{n,k,n',k'} \delta^{ab} \tr ( \chi^\dagger_{n,k}  [ (\Phi^\dagger_a)_{,n',k'}, (\Phi_b)_{,n+n',k+k'}] ) \, ,
\end{equation}
\begin{equation}
\label{T3}
T_3=\sum_{n,k,n',k'=0}^\infty P^{(0,2)}_{n,k,n',k'} \delta^{ab} \tr ( \chi^\dagger_{n,k}  \{ (\zeta^\dagger_{a})_{,n',k'}, (\zeta_b)_{n+n',k+k'} \} ) \, ,
\end{equation}
\begin{equation}
\label{T4}
T_4=\sum_{n,k,n',k'=0}^\infty P^{(0,3)}_{n,k,n',k'} \tr ( \chi^\dagger_{n,k} [ A^\dagger_{n',k'}, A_{n+n',k+k'}] ) \, .
\end{equation}
\begin{equation}
\label{T5}
T_5 = \frac{1}{2} \sum_{n,k,n',k'=0}^\infty P^{(1,1)}_{n,k,n',k'} \epsilon^{abc}  \tr \big( [ (\Phi^\dagger_a)_{n,k}, (\Phi^\dagger_b)_{n',k'}]  \zeta_{c,n+n',k+k'}  \big) \, ,
\end{equation}
\begin{equation}
\label{T6}
T_6 = \sum_{n,k,n',k'=0}^\infty P^{(1,2)}_{n,k,n',k'}
\delta^{ab}
\tr \big( [(\Phi^\dagger_a)_{n,k}, (\zeta_b)_{n',k'}^\dagger] A_{n+n',k+k'} \big) \, .
\end{equation}
One can be easily convinced that these are the correct objects.
When $i=0,$ the other raising operator $\tilde{V}$ can be any field with $j=0,1,2,3$, thus leaving the opportunity to define the four cubic generators \eqref{T1}--\eqref{T4}.
The Cartan generators $R_0, \tilde{R}_0$ have vanishing charge only if the fields are chosen to be $\tilde{V}=\hat{V}.$
The invariance under the other generators of $\SU(3)$ R-symmetry is achieved by building singlet structures.
While this is trivial for terms involving the fermion $\chi$ and the gauge field $A,$ instead the invariants involving the triplet scalars or fermions are built using the Kronecker delta.

In the case where the field $V$ transforms in the representation with $i=1,$  there are three possibilities for the field $\tilde{V}$, given by $j=0,1,2.$
The first case corresponds to the cubic generator $T_2$ defined in eq.~\eqref{T2}, while the new possibilities are \eqref{T5} and \eqref{T6}.
Notice that the invariance under $\SU(3)$ is achieved by using the two invariant objects at our disposal, {\sl i.e.}, the Levi-Civita symbol and the Kronecker delta.
When the field $V$ transforms in the representation $i=2,$, we have the two possibilities with $\tilde{V}$ transforming with $j=0,1.$
However, these cases corresponds to the generators \eqref{T3} and \eqref{T6}, respectively.
Similarly, when $i=3$ we can only choose $j=0,$ which is the case studied in eq.~\eqref{T4}.
Thus we conclude that there are six independent cubic structures in the fields which are separately invariant under the maximal bosonic subgroup $\SU(1,2) \times \SU(3) .$

\subsubsection*{Invariance under the fermionic generators}

According to the discussion on the invariance under the bosonic subgroup, we found that the most general cubic generator with the appropriate index structure is given by the linear combination \eqref{eq:generic_linear_combination_seccubic} with $A=1, \dots,6$.
Therefore, we reduced the ambiguity in the result to the determination of the real coefficients in the linear combination.
This will be uniquely fixed by requiring the invariance under all the fermionic generators of the PSU(1,2|3) spin group.

We consider the action of the supercharge $Q^{\dagger}_{4-a}$ on the 
terms \eqref{T1}--\eqref{T6}. We find
\begin{align}
	\begin{split}
 \lbrace Q^{\dagger}_{4-a}, T_1 \rbrace_D &= \frac{1}{2}
 \sum_{n,k,n',k'=0}^{\infty} \sqrt{n+n'+1} \,  P^{(0,0)}_{n,k;n'+1,k'}
\tr \le (\Phi_a)_{n+n',k+k'} \lbrace \chi^{\dagger}_{n,k} , \chi^{\dagger}_{n'+1,k'} \rbrace \ri  \, ,
\\
\lbrace Q^{\dagger}_{4-a}, T_2 \rbrace_D & = 
 \sum_{n,k,n',k'=0}^{\infty} 
\sqrt{n'+1}  P^{(0,1)}_{n,k;n'k'} 
\tr \le \chi^{\dagger}_{n'+1,k'} [(\Phi_a)_{n+n',k+k'} , \chi^{\dagger}_{n,k}] \ri \\
&  + \sum_{n,k,n',k'=0}^{\infty} \sqrt{n+n'+1} \,  P^{(0,1)}_{n,k;n'+1,k'} \,
\epsilon^{abc}
\tr \le (\zeta_c)_{n+n',k+k'} [ \chi^{\dagger}_{n,k} , (\Phi^{\dagger}_b)_{n'+1,k'} ] \ri  \, ,
\\
\lbrace Q^{\dagger}_{4-a}, T_3 \rbrace_D & =
 - \sum_{n,k,n',k'=0}^{\infty} \sqrt{n'+1} \, P^{(0,2)}_{n,k;n'k'} 
 \, \epsilon^{abc} 
 \tr \le (\Phi^{\dagger}_b)_{n'+1,k'} \lbrace (\zeta_c)_{n+n',k+k'} , \chi^{\dagger}_{n,k} \rbrace  \ri \\
&  + \sum_{n,k,n',k'=0}^{\infty} \sqrt{n+n'+1} \,  P^{(0,2)}_{n,k;n'+1,k'}
\tr \le A_{n+n',k+k'} \lbrace (\zeta^{\dagger}_a)_{n'+1,k'} , \chi^{\dagger}_{n,k} \rbrace \ri  \, ,
\\
\lbrace Q^{\dagger}_{4-a}, T_4 \rbrace_D& = 
 \sum_{n,k,n',k'=0}^{\infty} \sqrt{n'+1} \,  P^{(0,3)}_{n,k;n'k'}
\tr \le (\zeta^{\dagger}_a)_{n'+1,k'} [ A_{n+n',k+k'} , \chi^{\dagger}_{n,k} ] \ri  \, ,
\\
\lbrace Q^{\dagger}_{4-a}, T_5 \rbrace_D & =
 \sum_{n,k,n',k'=0}^{\infty} \sqrt{n} \, P^{(1,1)}_{n-1,k;n'+1,k'} \, 
\epsilon^{abc} \, 
\tr \le \chi^{\dagger}_{n,k} [(\Phi^{\dagger}_b)_{n'+1,k'} , (\zeta_c)_{n+n',k+k'}]  \ri  \\
&  + \frac{1}{2} \sum_{n,k,n',k'=0}^{\infty} \sqrt{n+n'+1} \, P^{(1,1)}_{n,k;n'+1,k'} \, 
\epsilon^{abc} \, 
\tr \le A_{n+n',k+k'} [(\Phi^{\dagger}_b)_{n,k} , (\Phi^{\dagger}_c)_{n'+1,k'}]  \ri  \, ,
\\
 \lbrace Q^{\dagger}_{4-a}, T_6 \rbrace_D & =
  \sum_{n,k,n',k'=0}^{\infty} \sqrt{n} \, P^{(1,2)}_{n'+1,k';n-1,k} \, 
\tr \le \chi^{\dagger}_{n,k} [(\zeta^{\dagger}_a)_{n'+1,k'} , A_{n+n',k+k'}]  \ri  \\
&   + \sum_{n,k,n',k'=0}^{\infty} \sqrt{n'+1} \, P^{(1,2)}_{n,k;n'k'} \, \epsilon^{abc} \,
\tr \le (\Phi^{\dagger}_c)_{n'+1,k'} [A_{n+n',k+k'} , (\Phi^{\dagger}_c)_{n,k}]  \ri  \, .
\end{split}
\end{align}
One can check that the following linear combination
\beq
\mathcal{Q} \equiv  T_1 + T_2 + T_3 + T_4 + T_5 - T_6 
\label{eq:linear_combination_fermgen}
\eeq
is invariant under the considered supercharge, {\sl i.e.}, it satisfies the condition
\beq
 \lbrace Q^{\dagger}_{4-a}, \mathcal{Q} \rbrace_D = 0 \, . 
\label{eq:vanishing_SUSY_variation_fermgenQdag}
\eeq
In order to show this result, we need to use the properties \eqref{eq:useful_identities_P-Qsupercharge} or the antisymmetry of the summations.
As a representative example, we show the vanishing of the following combination entering the Dirac brackets of the cubic generators \eqref{T1} and \eqref{T2}:
\beq
\begin{aligned}
&  \lbrace Q^{\dagger}_{4-a}, T_1 \rbrace_D 
+  \lbrace Q^{\dagger}_{4-a}, T_2 \rbrace_D = \\  
& =  \sum_{n,k,n',k'=0}^{\infty}  P^{(0,0)}_{n,k;n'+1,k'}
\le  \frac{1}{2} \sqrt{n+n'+1} - \frac{n'+1}{\sqrt{n+n'+1}} \ri
\tr \le (\Phi_a)_{n+n',k+k'} \lbrace \chi^{\dagger}_{n,k} , \chi^{\dagger}_{n'+1,k'} \rbrace \ri = \\
& =  \sum_{n,k,n',k'=0}^{\infty}  P^{(0,0)}_{n+1,k;n'+1,k'}
\frac{n-n'}{2 \sqrt{n+n'+2}} \, 
\tr \le (\Phi_a)_{n+n'+1,k+k'} \lbrace \chi^{\dagger}_{n+1,k} , \chi^{\dagger}_{n'+1,k'} \rbrace \ri = 0 \, .
\end{aligned}
\eeq
In going from the first to the second line we used the cyclicity properties of the trace and the identities \eqref{eq:useful_identities_P-Qsupercharge}, while in moving from the second to the third line we performed the summation and we shifted the label $n\rightarrow n+1$.
The last step is a consequence of the fact that we have a summation over $(n,n')$ of an odd expression in these indices.

%\vskip 5mm

%\noindent
We show in Appendix \ref{app:ferm_generators} that the linear combination \eqref{eq:linear_combination_fermgen} is also invariant under the action of the supercharges $Q_{4-a},$ namely
\beq
 \lbrace Q_{4-a}, \mathcal{Q} \rbrace_D = 0 \, . 
\label{eq:vanishing_SUSY_variation_fermgenQ}
\eeq
The identities \eqref{eq:vanishing_SUSY_variation_fermgenQdag} and \eqref{eq:vanishing_SUSY_variation_fermgenQ} are sufficient to prove the invariance under all the fermionic generators.
The proof proceeds by applying the graded Jacobi identity. 
Indeed, we consider 
\beq
 \lbrace L_+, \lbrace Q_{4-a}, \mathcal{Q} \rbrace_D \rbrace_D
 + \lbrace Q_{4-a}, \lbrace \mathcal{Q}, L_+ \rbrace_D  \rbrace_D
 - \lbrace \mathcal{Q} , \lbrace L_+, Q_{4-a} \rbrace_D \rbrace_D = 0 \, .
\eeq
Since we have proven before that the cubic generator $\mathcal{Q}$ is invariant under all the bosonic symmetries and under the action of $Q_{4-a},$ we have
\beq
\lbrace Q_{4-a}, \mathcal{Q} \rbrace_D = 
\lbrace \mathcal{Q}, L_+ \rbrace_D  = 0 \, .
\eeq
Now we use the commutation relation $\lbrace Q_{4-a},L_+ \rbrace_D =S^{\dagger}_{4-a}$ to conclude that
\beq
\lbrace S^{\dagger}_{4-a}, \mathcal{Q} \rbrace_D = 0 \, .
\eeq
We apply a similar trick by considering the graded Jacobi identity 
\beq
\lbrace J_+, \lbrace Q_{4-a}, \mathcal{Q} \rbrace_D \rbrace_D
 + \lbrace Q_{4-a}, \lbrace \mathcal{Q}, J_+ \rbrace_D  \rbrace_D
 - \lbrace \mathcal{Q} , \lbrace J_+, Q_{4-a} \rbrace_D \rbrace_D = 0 \, ,
\eeq
plus the commutation relation $\lbrace Q_{4-a}, J_+ \rbrace_D =\tilde{Q}_{4-a}$.
This allows to conclude that
\beq
\lbrace \tilde{Q}_{4-a}, \mathcal{Q} \rbrace_D = 0 \, .
\eeq
One can work in the same way by starting from the result $\lbrace Q^{\dagger}_{4-a}, \mathcal{Q} \rbrace_D = 0,$ to derive that 
\beq
\lbrace \tilde{Q}^{\dagger}_{4-a}, \mathcal{Q} \rbrace_D = \lbrace S_{4-a}, \mathcal{Q} \rbrace_D = 0 \, .
\eeq
This shows that the linear combination \eqref{eq:linear_combination_fermgen} is invariant under all the fermionic generators of $\mathrm{PSU}(1,2|3)$ group.

\subsubsection*{Reduction to subsectors}

The invariance under supersymmetry uniquely fixed the relative coefficients in the linear combination \eqref{eq:linear_combination_fermgen}.
A consistency check of this result comes from the reduction of the general expression to the PSU(1,1|2) subsector, where the method described in this Section was originally applied \cite{Beisert:2007sk, Zwiebel:2007cpa,Baiguera:2021hky}.
The restriction to this case can be achieved by setting some of the fields to zero
\begin{equation}
\Phi_3 = 0 \, , \qquad A = 0 \, , \qquad \zeta_1 = \zeta_2 =0 \, .
\end{equation}
Furthermore, we set $k=0,$ specifying the dictionary 
\begin{equation}
(\Phi_1)_{n}= (\Phi_1)_{n,0} \, , \qquad
 (\Phi_2)_{n}= (\Phi_2)_{n,0} \, , \qquad
  (\psi_1)_{n}= - (\zeta_3)_{n,0} \, , \qquad
  (\psi_2)_n=  \chi_{n+1,0}  \, ,
   \label{eq:psu112_dictionary}
\end{equation}
where the fields on the left-hand side refer to the notation used in \cite{Baiguera:2021hky}, while the fields on the right-hand side refer to the notation used in this work. 
In this way, the cubic supercharge $\mathcal{Q}$ reduce to
\begin{equation}
\mathcal{Q} = T_1 + T_2 + T_3 + T_5 \, .
\end{equation}
with the cubic generators being
\begin{equation}
T_1= \frac{1}{2} \sum_{n,n'=0}^\infty \sqrt{\frac{n+n'+2}{(n+1)(n'+1)}} \tr ( (\psi^\dagger_2)_{n} \{ (\psi^\dagger_2)_{n'}, (\psi_2)_{n+n'+1} \} ) \, ,
\end{equation}
\begin{equation}
T_2=\sum_{n,n'=0}^\infty \frac{1}{\sqrt{n+1}} \delta^{ab} \tr ( (\psi^\dagger_2)_{n}  [ (\Phi^\dagger_a)_{n'}, (\Phi_b)_{n+n'+1}] ) \, ,
\end{equation}
\begin{equation}
T_3=\sum_{n,n'=0}^\infty \sqrt{\frac{n' + 1}{(n+n' +2)(n+1) }}  \tr ( (\psi^\dagger_2)_{n}  \{ (\psi^\dagger_1)_{n'}, (\psi_1)_{n+n'+1} \} ) \, ,
\end{equation}
\begin{equation}
T_5 = -  \sum_{n,n'=0}^\infty  \frac{1}{\sqrt{n+n' + 1}}   \tr \big( [ (\Phi^\dagger_1)_{n}, (\Phi^\dagger_2)_{n'}]  (\psi_1)_{n+n'}  \big) \, ,
\end{equation}
while $T_4=T_6=0$ since there is no dynamical gauge field in this subsector.
With this, the cubic supercharge $\mathcal{Q}$ is exactly the one  introduced in reference \cite{Baiguera:2021hky} for the PSU(1,1|2) subsector.

One can perform a further reduction to the SU(1,1|1) subsector by setting
\begin{equation}
\Phi_2 = 0 \, \qquad \psi_1 = 0 \, ,
\end{equation} 
and renaming $\Phi_1=\Phi$ and $\psi_2=\psi$.
Then the supercharge is now reduced to $\mathcal{Q} = T_1 + T_2,$ which matches exactly that of 
the $\SU(1,1|1)$ case in \cite{Baiguera:2021hky}.

{\sl{En passant}}, we notice that it is not possible to define the cubic supercharge \eqref{eq:linear_combination_fermgen} in the SU(1,2|2) subsector, even if part of the supersymmetry of $\mathcal{N}=4$ SYM is preserved.
The reason is that the expressions \eqref{T1}--\eqref{T6} always contain at least one singlet fermion $\chi$.
Since this field is only non-vanishing in the SU(1,1|1) and PSU(1,1|2) subsectors, we conclude that the previous construction does not work in the other cases. 
It will instead be possible to recover SU(1,2|2) as a special case by setting the appropriate fields to zero at the level of the interacting Hamiltonian.

%%%%%%%%%%%%%%%%%
\subsection{Derivation of the interacting Hamiltonian}
\label{ssec:derivation_Hint_cubic}

Starting from the cubic supercharge \eqref{eq:linear_combination_fermgen}, we compute the interacting Hamiltonian by using the identity 
\beq
\lbrace \mathcal{Q} , \mathcal{Q}^{\dagger} \rbrace_D = H_{\rm int} \, .
\label{eq:sec_cubic_supercharge}
\eeq
It is a tedious but straightforward exercise to find all the terms arising from this Dirac bracket.
Remarkably, one can show that the interacting Hamiltonian can be organized in the compact form
\begin{align}\label{eq:Hint_tot_cubic}
	\begin{split}
		H_{\rm int} &= H_D + H_F \, ,
		\\
H_D &=	\sum_{n,k=0}^{\infty}
		\tr \left[ (\mathcal{B}_0^{\dagger})_{n,k} (\mathcal{B}_0)_{n,k}  + \sum_{a=1}^3 \sum_{I=1,2} (\mathcal{B}_I^{a\dagger})_{n,k} (\mathcal{B}_I^a)_{n,k}  + (\mathcal{B}_3^{\dagger})_{n,k} (\mathcal{B}_3)_{n,k} \right] \, ,  \\
H_F	&= \sum_{n,k=0}^{\infty}
		\tr \left[ (\mathcal{F}_0^{\dagger})_{n,k} (\mathcal{F}_0)_{n,k}  + \sum_{a=1}^3 \sum_{I=1,2} (\mathcal{F}_I^{a\dagger})_{n,k} (\mathcal{F}_I^a)_{n,k}  + (\mathcal{F}_3^{\dagger})_{n,k} (\mathcal{F}_3)_{n,k} \right]   \, ,
	\end{split}
\end{align}
where we introduce the following set of $\mathcal{B}$ blocks: 
\begin{align}
	(\mathcal{B}_0)_{n,k}  & = \sum_{n',k'=0}^{\infty} \le  P^{(0,0)}_{n,k;n',k'} \lbrace \chi^{\dagger}_{n',k'} , \chi_{n+n',k+k'} \rbrace +
	\sum_{a=1}^3  P^{(0,1)}_{n,k;n',k'} [(\Phi^{\dagger}_a)_{n',k'} , (\Phi_a)_{n+n',k+k'}] \right. \nonumber \\ 
	& \left.  + \sum_{a=1}^3 P^{(0,2)}_{n,k;n',k'} \lbrace (\zeta^{\dagger}_a)_{n',k'} , (\zeta_a)_{n+n',k+k'} \rbrace  +  P^{(0,3)}_{n,k;n',k'} [A^{\dagger}_{n',k'} , A_{n+n',k+k'}]   \ri \, ,
\label{eq:block_B0} \\  \nonumber
	(\mathcal{B}_{1}^a)_{n,k} & \equiv
	\sum_{n',k'=0}^{\infty} \le  P^{(1,1)}_{n,k:n',k'}   \epsilon^{abc} 
	[ (\zeta_b)_{n+n',k+k'} , (\Phi^{\dagger}_c)_{n', k'} ]  \right. \\
	& \left.  
	-  P^{(1,2)}_{n,k;n',k'} [ (\zeta^{\dagger}_a)_{n',k'}  , A_{n+n',k+k'} ] + P^{(1,0)}_{n,k;n',k'}  [ (\Phi_a)_{n+n',k+k'} , \chi^{\dagger}_{n',k'}  ]  \ri \, ,
\label{eq:block_B1} \\
	(\mathcal{B}_{2}^a)_{n,k} & \equiv 
	\sum_{n',k'=0}^{\infty}  \le  P^{(2,1)}_{n,k;n',k'}  [(\Phi^{\dagger}_a)_{n',k'}  , A_{n+n',k+k'}]
	  +  P^{(2,0)}_{n,k;n',k'}  \lbrace (\zeta_a)_{n+n',k+k'} , \chi^{\dagger}_{n',k'}    \rbrace  \ri \, ,
\label{eq:block_B2} \\
(\mathcal{B}_3)_{n,k} & \equiv \sum_{n',k'=0}^{\infty} P^{(3,0)}_{n,k;n',k'}  [ A_{n+n',k+k'} , \chi^{\dagger}_{n',k'}  ] \, .
\label{eq:block_B3}
\end{align}
We define the $\mathcal{F}_{n,k}$ blocks as
\begin{align}
		(\mathcal{F}_0)_{n,k} & \equiv \frac{1}{2} \sum_{n'=0}^{n}\sum_{k'=0}^k P^{(0,0)}_{n',k';n-n',k-k'} \{\chi_{n-n',k-k'},\chi_{n',k'} \} \, , 
		\label{eq:block_F0}  \\
		(\mathcal{F}_1^a)_{n,k} & \equiv \sum_{n'=0}^n \sum_{k'=0}^k P^{(0,1)}_{n',k',n-n',k-k'} [(\Phi_a)_{n-n',k-k'},\chi_{n',k'}] \, ,
		\label{eq:block_F1}  \\  \nonumber
		(\mathcal{F}_2^a)_{n,k} & \equiv \frac{1}{2}  \sum_{n'=0}^{n}\sum_{k'=0}^k P^{(1,1)}_{n',k',n-n',k-k'} \epsilon^{abc} 	[(\Phi_c)_{n',k'},(\Phi_b)_{n-n',k-k'}]\\
 &	 + P^{(0,2)}_{n',k',n-n',k-k'} \{(\zeta_a)_{n-n',k-k'},\chi_{n',k'} \} \, ,
		\label{eq:block_F2} \\ 
		(\mathcal{F}_3)_{n,k} & \equiv \sum_{n'=0}^{n}\sum_{k'=0}^k P^{(1,2)}_{n',k',n-n',k-k'} [(\zeta_a)_{n-n',k-k'},(\Phi_a)_{n',k'}]  - P^{(0,3)}_{n',k';n-n'k-k'} [A_{n-n',k-k'},\chi_{n',k'}] \, .
	\label{eq:block_F3}
\end{align}
The $\mathcal{B}$ blocks originate the D-terms $H_D$ in eq.~\eqref{eq:Hint_tot_cubic}, while the $\mathcal{F}$ blocks generate the contribution $H_F$ to the interacting Hamiltonian composed of F-terms.
The total Hamiltonian \eqref{eq:Hint_tot_cubic} comprises all the contributions of the PSU(1,2|3) sector.
This is the largest possible spin group admitting a near-BPS limit of the form \eqref{eq:SMT_limits} and contains all the other admissable cases as subsectors.
At the level of the cubic supercharge \eqref{eq:linear_combination_fermgen} there is the obstacle that we can only restrict to subsectors containing the singlet fermion $\chi,$ since it is essential to build all the structures \eqref{eq:generic_TA_term}.
Instead this problem does not occur when considering the full Hamiltonian: we can therefore recover all the results derived in the references \cite{Harmark:2019zkn,Baiguera:2020jgy,Baiguera:2020mgk,Baiguera:2021hky} by setting the corresponding fields to zero.

We comment more explicitly the classes of terms entering the result:
\begin{itemize}
\item \textbf{Charge density.}
The charge density 
\beq
\mathbf{Q}_{n,k}= \frac{1}{\sqrt{n+k}} (\mathcal{B}_0)_{n,k}
\label{eq:def_charge_density_cubicgen}
\eeq
corresponds to the block \eqref{eq:block_B0}.
It is a natural extension of the charge densities derived in \cite{Harmark:2019zkn,Baiguera:2020jgy,Baiguera:2020mgk,Baiguera:2021hky} for all the other subsectors of PSU(1,2|3).\footnote{In order to compare with the references \cite{Harmark:2019zkn,Baiguera:2020jgy,Baiguera:2020mgk,Baiguera:2021hky}, we point out that therein we used the notation $\mathbf{Q}_{n,k}$ for the charge densities, referring to lower-case $q$ to denote the separate contributions from each field. }
Terms of this kind arise when contracting two singlet fermions among the structures \eqref{T1}--\eqref{T4}.
\item \textbf{Generalization of SU(1,2|2) blocks.}
We define
\beq
(\mathcal{B}_{1}^a)_{n,k} \equiv (F_a - K_a + \mathcal{H}_a)_{n,k} \, , \qquad
(\mathcal{B}_{2}^a)_{n,k} \equiv (W_a + \mathcal{M}_a)_{n,k} \, ,
\label{eq:definition_FKHWM}
\eeq
where each structure refers to a corresponding term in the blocks introduced in eqs.~\eqref{eq:block_B1} and \eqref{eq:block_B2}.
The objects $F_a, K_a, W_a$ generalize the homonymous blocks appearing in the Hamiltonian (3.100) of the SU(1,2|2) subsector presented in reference \cite{Baiguera:2020mgk}.
This part of the interactions arises from contractions involving the blocks \eqref{T5} and \eqref{T6} only.
\item \textbf{Generalization of PSU(1,1|2) F-terms.}
The F-blocks in eq.~\eqref{eq:block_F0}--\eqref{eq:block_F3} are a generalization of all the interactions (besides the charge density term) entering the PSU(1,1|2) subsector, see the Hamiltonian (3.56) in reference \cite{Baiguera:2021hky}.
Here we observe one of the advantages of the cubic supercharge method: the organization of the structure \eqref{eq:linear_combination_fermgen} naturally separates D-terms and F-terms, while the spherical expansion procedure applied in \cite{Baiguera:2021hky} gave a structure which highlighted the symmetry between the fermionic fields therein called as $\psi_1, \psi_2.$
By applying the dictionary in eqs.~\eqref{eq:psu112_dictionary} and \eqref{eq:psu112_dictionary}, one can check that the full Hamiltonian \eqref{eq:Hint_tot_cubic} indeed reduces to the one of the PSU(1,1|2) subsector.
\item \textbf{New D-terms.}
There are new D-term structures, entering eqs.~\eqref{eq:block_B1}--\eqref{eq:block_B3}, that arise from the brackets involving the cubic generators in eqs.~\eqref{T1}--\eqref{T4}. 
They come from contractions where a singlet fermion (and its hermitian conjugate) survive.
\item \textbf{New F-terms.}
Finally, we have one additional F-term which involves the gauge field. It is the last contribution of the block $\mathcal{F}_3$ in eq.~\eqref{eq:block_F3}.
\end{itemize}

\noindent
According to the identity \eqref{eq:sec_cubic_supercharge}, the Hamiltonian was obtained as the anticommutator of a complex supercharge and its hermitian conjugate.
Standard manipulations involving supersymmetry show that the spectrum of such a theory is positive definite \cite{Bilal:2001nv}.
The interactions written as in eq.~\eqref{eq:Hint_tot_cubic} confirm this argument: they appear in a form which manifestly shows the positivity of each term, since they are written as a product of a block times its hermitian conjugate.
We will see in Section \ref{sec:symmetry_structure} that these blocks provide a convenient way to show the symmetry structure of the result.

%%%%%%%%%%%%%%%%%%%%%
\section{Symmetry structure of the Hamiltonian}
\label{sec:symmetry_structure}

\subsection{Symmetry structure of D-terms and F-terms}
\label{ssec:actions_supercharges}
In this Section, we discuss the symmetry structure of the Hamiltonian \eqref{eq:Hint_tot_cubic}. 
As expected, the analysis in the following will show that the Hamiltonian is manifestly invariant under the PSU$(1,2|3)$ symmetry. However, as we shall see, this is also the case for the separate D-term and F-term parts of the Hamiltonian. 

Let us focus on the bosonic part of the symmetry group first, whose action on the letters was summarized in eq.~\eqref{eq:SU122-block-I}. 
We collect the blocks in the following way
\begin{equation}
	W_I = \{\mathcal{B}_I\,, \mathcal{F}_I\}, \qquad I=0,1,2,3
\end{equation}
where the SU$(3)$ indices $a$ for $I=1,2$ are not displayed explicitly.
Then the crucial observation is that $\mathcal{B}_I$ and $\mathcal{F}_I$ making up the Hamiltonian \eqref{eq:Hint_tot_cubic} exactly transform in the $(p,q)=(0,I-3)$ representations of the SU$(1,2)$ algebra for $I=0,1,2,3$, {\sl{i.e.}}, the equations \eqref{eq:SU122-block-I} are also satisfied by the blocks $W_I$.
This is the generalization of the example already shown in \cite{Baiguera:2020mgk}, which was the SU$(1,2|2)$ subsector written in terms of $\mN=2$ vector multiplets, where the letters transform in $I=1,2,3$ representations, while the $\mathcal{B}_I$ blocks transform in $I=0,1,2$ representations.  

In conclusion, the blocks $\mathcal{B}_I$ and $\mathcal{F}_I$ transform as irreducible representations of both SU$(1,2)$ as well as SU$(3)$. When combining this into $H_D$ and $H_F$ using \eqref{eq:Hint_tot_cubic}, this shows that $H_D$ and $H_F$ are both invariant under the SU$(1,2) \times \mbox{SU}(3)$ bosonic symmetry transformations. 

\begin{figure}
	\centering
	\includegraphics[trim=3cm 12cm 10cm 4cm,width=0.35\linewidth]{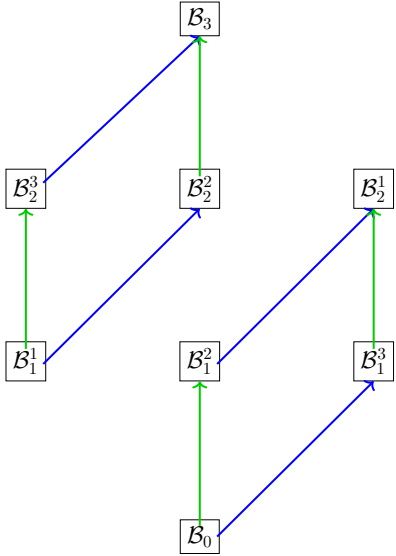}
	\caption{Action by supercharges on the blocks $\mathcal{B}.$
	The blue lines represent the transformation $Q_1^\dagger$, while the green lines are the transformation of supercharge $S_2$. The actions on the $\mathcal{F}$ blocks follow in the same way. }
	\label{fig:block123super}
\end{figure}

Next, let us move on to the supersymmetry.
We find the Hamiltonian $H_D$ and $H_F$ are {\sl{separately}} invariant under the action of all supercharges $Q_a, \tilde{Q}_a$ and $S_a$. 
There are some basic patterns followed by the action of supercharges. 
First of all, due to the SU$(3)_R$ symmetry of $Q_a$ supercharges \eqref{eq:R-symmetry-letters}, for different $a=1,2,3$, the supercharges are relating different blocks $\mathcal{W}_I^a$ as $\mN=1$ chiral multiplets or $\mN=1$ vector multiplets. 
On the other hand, we notice from the letter representation \eqref{eq:supercharge-asletters} that the supercharges $Q_a$ act on the descendants generated by $d_1$ letters, while $\tilde{Q}_a$ act on the descendants generated by $d_2$. 
This can be seen from the fact that their actions shift the levels of $n$ or $k$ by $1$, which counts the descendant levels of $d_1,d_2$, respectively. 
The $S_a$ supercharges act on both directions simultaneously. 
In total, the difference between the three classes of supercharges are majorly reflected at the level of descendants, which are also closely related to the three different components of momenta saturated Clebsch-Gordan coefficients.\footnote{The CG coefficients in eq.~\eqref{eq:relation_P_CG} present three different combinations of momenta satisfying a precise saturation. These labels correspond to the action of the three classes of supercharges.} 
We will also discuss this point in appendix \ref{app:details_invariance_cubic}. 
Without loss of generality, let's consider the action by $Q_1^\dagger$ as an example. 
The definition of this supercharge can be found in eq.~\eqref{eq:supercharge-asletters}.
We can then check that the blocks transform as supermultiplets under the actions of supercharges.
\begin{align}
	\begin{split}
		\{Q_1^\dagger, (\mathcal{B}_0)_{n,k} \}_D &= \sqrt{n+1} (\mathcal{B}_1^3)_{n,k}, \qquad \{Q_1^\dagger, (\mathcal{B}_0^\dagger)_{n,k} \}_D = 0 \\
		\{Q_1^\dagger, (\mathcal{B}_1^1)_{n,k} \}_D &= \sqrt{n} (\mathcal{B}_2^2)_{n-1,k}, \qquad \{Q_1^\dagger, (\mathcal{B}_1^{1\dagger} )_{n,k}\}_D = 0 \\
		\{Q_1^\dagger, (\mathcal{B}_1^2)_{n,k} \}_D &= - \sqrt{n} (\mathcal{B}_2^1)_{n-1,k}, \qquad \{Q_1^\dagger, (\mathcal{B}_1^{2\dagger} )_{n,k}\}_D = 0  \\
		\{Q_1^\dagger, (\mathcal{B}_1^3)_{n,k} \}_D &= 0, \qquad \{Q_1^\dagger, (\mathcal{B}_1^{3\dagger} )_{n,k}\}_D = -\sqrt{n+1} (\mathcal{B}_0^\dagger)_{n+1,k} \\
		\{Q_1^\dagger, (\mathcal{B}_2^1)_{n,k} \}_D &= 0, \qquad \{Q_1^\dagger, (\mathcal{B}_2^{1\dagger} )_{n,k}\}_D =  -\sqrt{n+1} (\mathcal{B}_1^{2\dagger})_{n+1,k} \\
		\{Q_1^\dagger, (\mathcal{B}_2^2)_{n,k} \}_D &= 0, \qquad \{Q_1^\dagger, (\mathcal{B}_2^{2\dagger} )_{n,k}\}_D =  \sqrt{n+1} (\mathcal{B}_1^{1\dagger})_{n+1,k} \\
		\{Q_1^\dagger, (\mathcal{B}_2^3)_{n,k} \}_D &= \sqrt{n} (\mathcal{B}_3)_{n-1,k}, \qquad \{Q_1^\dagger, (\mathcal{B}_2^{3\dagger} )_{n,k}\}_D =  0 \\
		\{Q_1^\dagger, (\mathcal{B}_3)_{n,k} \}_D &= 0, \qquad \{Q_1^\dagger, (\mathcal{B}_3^{\dagger} )_{n,k}\}_D =  -\sqrt{n+1} (\mathcal{B}_2^{3\dagger})_{n+1,k}
	\end{split}
\end{align}
We also compute 
\begin{align}
	\begin{split}
		& \{Q_1^\dagger, (\mathcal{F}_0)_{n,k} \}_D = \sqrt{n} (\mathcal{F}_1^3)_{n-1,k}, \qquad \{Q_1^\dagger, (\mathcal{F}_0^\dagger)_{n,k} \}_D = 0 \\
		& \{Q_1^\dagger, (\mathcal{F}_1^1)_{n,k} \}_D = \sqrt{n} (\mathcal{F}_2^2)_{n-1,k}, \qquad \{Q_1^\dagger, (\mathcal{F}_1^{1\dagger})_{n,k} \}_D = 0  \\
		& \{Q_1^\dagger, (\mathcal{F}_1^2)_{n,k} \}_D = -\sqrt{n} (\mathcal{F}_2^1)_{n-1,k}, \qquad \{Q_1^\dagger, (\mathcal{F}_1^{2\dagger})_{n,k} \}_D = 0 \\
		& \{Q_1^\dagger, (\mathcal{F}_1^3)_{n,k} \}_D = 0, \qquad \{Q_1^\dagger, (\mathcal{F}_1^{3\dagger})_{n,k} \}_D = -\sqrt{n+1} (\mathcal{F}_0^\dagger)_{n+1,k}
		\\
		& \{Q_1^\dagger, (\mathcal{F}_2^1)_{n,k} \}_D = 0, \qquad \{Q_1^\dagger, (\mathcal{F}_2^{1\dagger})_{n,k} \}_D = -\sqrt{n+1} (\mathcal{F}_1^{2\dagger})_{n+1,k} \\
		& \{Q_1^\dagger, (\mathcal{F}_2^2)_{n,k} \}_D = 0, \qquad \{Q_1^\dagger, (\mathcal{F}_2^{2\dagger})_{n,k} \}_D = \sqrt{n+1} (\mathcal{F}_1^{1\dagger})_{n+1,k} \\
		& \{Q_1^\dagger, (\mathcal{F}_2^3)_{n,k} \}_D = -\sqrt{n} (\mathcal{F}_3)_{n-1,k}, \qquad \{Q_1^\dagger, (\mathcal{F}_2^{3\dagger})_{n,k} \}_D = 0 \\
		&  \{Q_1^\dagger, (\mathcal{F}_3)_{n,k} \}_D = 0, \qquad \{Q_1^\dagger, (\mathcal{F}_3^{\dagger})_{n,k} \}_D = \sqrt{n+1} (\mathcal{F}_2^{3\dagger})_{n+1,k}
	\end{split}
\end{align}
Then in total 
\begin{equation}
	\{Q_1^\dagger ,H_D\}_D= \{Q_1^\dagger,H_F \}_D=0
	\label{eq:invariance_DF_terms}
\end{equation}
The action of the supercharges for the blocks $\mathcal{B}_I$ is graphically shown in Figure \ref{fig:block123super}. 
Using that $H_D$ and $H_F$ are separately invariant under SU$(1,2) \times \mbox{SU}(3)$, one can argue using the Jacobi-identity, that it follows from invariance under $Q_1^\dagger$ that they are separately invariant under all the supercharges $Q_a^\dagger$, $\tilde{Q}_a^\dagger$ and $S_a^\dagger$. Since $H_D$ and $H_F$ also are hermitian, it follows they are also separately invariant under all the supercharges $Q_a$, $\tilde{Q}_a$ and $S_a$.

Therefore, any Hamiltonian of the form  $H_D+\Lambda H_F$ is invariant under the PSU$(1,2|3)$ global symmetry action, {\sl{i.e.}}, there is a free coefficient $\Lambda$ that does not spoil the invariance.
The Hamiltonian $H_D+ H_F$ obtained from the decoupling limit of $\mN=4$ SYM might indicate an the existence of a further enhanced symmetry that fixes $\Lambda =1$, as we will discuss below. 

Furthermore, it is important to remark that the fact that both the $\mathcal{B}$ and $\mathcal{F}$ blocks transforms well under the full PSU$(1,2|3)$ algebra, can be used to cast $H_D$ and $H_F$ in \eqref{eq:Hint_tot_cubic} separately as norms in the representation space of PSU$(1,2|3)$ as in the construction of \cite{Baiguera:2020mgk}. This shows the origin of our findings in \cite{Baiguera:2020mgk} for the SU$(1,1|1)$ and SU$(1,2|2)$ Spin Matrix theories, here with the remarkable extension to the full PSU$(1,2|3)$ symmetry, and to both the D-terms and F-terms separately.

Both the $\mathcal{B}$ and $\mathcal{F}$ blocks can be organized into three $\mN=1$ chiral multiplets and $\mN=1$ vector multiplets. 
Thus the invariance of the Hamiltonian is manifest, since both terms are made by an $\mN=3$ vector multiplet. 
For a discussion of field theories with $\mN=3$ supersymmetry, see for example  \cite{Argyres:2019yyb}. 
The CPT invariance of lagrangian theories implies the $\mN=3$ theories get enhanced to $\mN=4$ supersymmetric gauge theories \cite{DHoker:1999yni}. 
The situation in the current case could be more subtle.
First of all, the field theory description of PSU$(1,2|3)$ is less clear. 
The global symmetry SU$(1,2)$ is neither the conformal group nor includes the Lorentz symmetry.
Secondly, the programme of formulating the local field description of SMT was initiated in \cite{Baiguera:2020jgy}, where we found that the field theory description of SU$(1,1)$ subsector is both semi-local and ghost-like.\footnote{By semi-local, we mean that the Dirac brackets satisfied by the fields are not Dirac delta functions, but their Fourier representation where the summation is only performed over positive modes. } 
This behaviour could be potentially related to the chiral algebra \cite{Beem:2013sza} of $4d$ SCFT. 
Its generalization to SU$(1,2)$ subsectors could be more non-trivial. 
Besides, due to the non-relativistic nature of Spin matrix theory, there are no anti-particle excitations in the theory. 
Due to many novel features of the field theory describing the Spin matrix theory, whether the $\mN=3$ supersymmetry gets enhanced to $\mN=4$ supersymmetry in the current model should be analysed more systematically. 
We will leave this issue as topic for a future work.

The explicit formula of the PSU$(1,2|3)$ Hamiltonian \eqref{eq:Hint_tot_cubic} enables us to improve our understanding of other subsectors. 
As remarked in section \ref{ssec:derivation_Hint_cubic}, we can now resolve the puzzles raised in the PSU$(1,1|2)$ subsector \cite{Baiguera:2021hky}.
From the spin chain point of view \cite{Bellucci:2006bv}, it is not manifest that the Hamiltonian is positive definite. 
However, once we take the decoupling limit of the PSU$(1,2|3)$ Hamiltonian to acquire the PSU$(1,1|2)$ subsector, the corresponding Hamiltonian is made by a D-term $\mN=2$ hypermultiplet and an F-term $\mN=2$ hypermultiplet. 
Both the manifest invariance under PSU$(1,1|2)$ global symmetry and the positive definiteness are ensured by the latter formulation. 
This also explains why the SU$(2)$ automorphism between fermions in PSU$(1,1|2)$ subsector is ``emergent" \cite{Beisert:2007sk},
as there is no automorphism in PSU$(1,2|3)$ sector which can act on fermion singlet $\chi$ and fermion triplet $\zeta^a$.

\subsection{Uniqueness of the Hamiltonian}
\label{ssec:uniqueness_ham}

Now we argue that the Hamiltonian \eqref{eq:Hint_tot_cubic} derived using the cubic supercharge method \eqref{eq:sec_cubic_supercharge} is unique.
The prescription given at the beginning of Section \ref{ssec:construction_cubic_supercharge} to build the fermionic generator $\mathcal{Q}$ assumed that each term $T_A$ entering the linear combination \eqref{eq:generic_linear_combination_seccubic} was independently invariant under the full bosonic subgroup of PSU(1,2|3), while the fermionic part of the spin group is used to restrict the coefficients $\alpha_A$.
However, one may in principle choose a more general set of cubic terms in the fields such that their linear combination  (anti)commutes with the bosonic generators, while they separately do not. 

A way to fix this ambiguity consists in comparing the result with other techniques.
In Section \ref{ssec:bosonic_interactions_sphere_red} we will derive the purely bosonic part of the effective Hamiltonian using the spherical expansion procedure (method 2 of the list presented in Section \ref{sec:introduction}).
In particular, we will show that the bosonic interactions derived with the spherical expansion precisely match with the result \eqref{eq:Hint_tot_cubic}.
These terms involving the interactions between scalars and the bosonic gauge fields, and we dub it as $H_{\text{bos}}$.  
This partial comparison is sufficient to argue that the full Hamiltonian, including the fermionic terms, matches between the two procedures.
For the sake of argument, we assume that there exist two different Hamiltonians with the same bosonic interactions $H_{\rm bos},$ but different fermionic ones:
\beq
H_{\rm int,1} = H_{\rm bos} + H_{\rm ferm,1} \, , \qquad
H_{\rm int,2} = H_{\rm bos} + H_{\rm ferm,2} \, .
\label{eq:splitting_sec42}
\eeq
In these expressions, we denoted with $H_{\rm bos}$ the terms containing purely bosonic fundamental fields, and with $H_{\rm ferm}$ the remaining parts, which collect together the purely fermionic interactions and the mixed terms having both bosons and fermions.\footnote{It should be noted that in lower-dimensional field theories, bosonization/fermionization dualities can relate fields with different statistics,  {\sl{e.g.}}, \cite{gogolin1999bosonization,senechal2004introduction,Benini:2017aed}, which might also extend to 4d quantum field theory \cite{Furusawa:2018izz,Murugan:2021jwu}.
Therefore, one may be worried that the distinction between bosonic and fermionic interactions is ambiguous.
However, fields in the SMT Hamiltonian surviving the near-BPS limits \eqref{eq:SMT_limits} arise from fundamental fields defined in the original $\mathcal{N}=4$ SYM action, where bosons and fermions obey commutative and anti-commutative Dirac brackets respectively, see eqs.~\eqref{eq:Dirac_brackets_singl_ferm}--\eqref{eq:Dirac_brackets_gauge}. }

The splitting \eqref{eq:splitting_sec42} must be satisfied by any possible near-BPS Hamiltonian, since we assume that the purely bosonic part is unambiguously fixed with the spherical expansion method.
We then find that by taking the difference between the two expressions, we produce another interacting Hamiltonian that does not contain any bosonic part
\beq
H_{\rm int,new} = H_{\rm ferm,1}  - H_{\rm ferm,2} \, . 
\eeq
A near-BPS effective Hamiltonian without purely bosonic terms cannot be invariant under the full symmetry group PSU(1,2|3), in particular supersymmetry invariance would necessarily be violated.
This can be checked explicitly by applying the SUSY transformations induced by the generators \eqref{eq:supercharge-asletters} to any expression without bosonic letters.
Therefore, we conclude that it was not possible to have two Hamiltonians with different fermionic interactions, thus showing the uniqueness of the construction.

Notice that the previous argument does not make any requirement on the splitting of the interacting Hamiltonian into D- and F-terms, that we observed in Section \ref{ssec:actions_supercharges} to be separately invariant under the full PSU(1,2|3) symmetry group.
Indeed, the purely bosonic Hamiltonian $H_{\rm bos}$ will itself contain a D-term and an F-term, and the same will be true for $H_{\rm ferm}.$
However, this sub-structure can be ignored for the purposes of the previous derivation, without affecting the final conclusion.
Having shown that the Hamiltonian is unique, this procedure also fixes unambiguously the free coefficient $\Lambda$ mentioned below eq.~\eqref{eq:invariance_DF_terms} to be 1.

%%%%%%%%%%%%%%%%%%%%%
\section{Hamiltonian from spherical expansion}
\label{sec:ham_sphere}

In this Section we apply the spherical expansion method (bullet 2 in Section \ref{sec:introduction}) to compute the effective Hamiltonian of the PSU(1,2|3) SMT.
This provides an alternative procedure to derive the interactions \eqref{eq:Hint_tot_cubic}, but has the advantage to completely fix the purely bosonic part of the Hamiltonian, which is essential to prove the uniqueness of the result (see discussion in Section \ref{ssec:uniqueness_ham}.

The outline of the Section is the following.
We begin by reviewing the general setting to perform the spherical expansion in Section \ref{ssec:general_procedure_sphere_red}.
Then we derive the free part of the Hamiltonian and the interactions in the purely bosonic sector in Sections \ref{ssec:free_Ham_sphere_red} and \ref{ssec:bosonic_interactions_sphere_red}, respectively.
We finally compare to the results obtained using the cubic supercharge technique in Section \ref{ssec:comparison_cubic_supercharge}.

\subsection{General procedure}
\label{ssec:general_procedure_sphere_red}

We set the conventions for the spherical expansion of the classical $\mathcal{N}=4$ SYM theory defined on $\mathbb{R} \times S^3$ by following the same notation as in references \cite{Harmark:2019zkn,Baiguera:2020jgy,Baiguera:2020mgk,Baiguera:2021hky}. We summarize them here, starting from the action
\begin{multline}
\label{eq:N4SYMaction}
S = \int_{\mathbb{R} \times S^3} \sqrt{-\mathrm{det} \, g_{\mu\nu}} \, \tr \left\lbrace - \frac{1}{4} F_{\mu\nu}^2 - |D_{\mu} \Phi_a|^2 - |\Phi_a|^2 - i \psi^{\dagger}_a \bar{\sigma}^{\mu} D_{\mu} \psi^A + g \sum_{A,B,a} C_{AB}^{a} \psi^A [\Phi_a, \psi^B]
  \right. \\
\left. + g \sum_{A,B,a} \bar{C}^{aAB} \psi^{\dagger}_A [\Phi^{\dagger}_a, \psi^{\dagger}_B]
- \frac{g^2}{2} \sum_{a,b} \le |[\Phi_a, \Phi_b]|^2 + |[\Phi_a, \Phi^{\dagger}_b]|^2 \ri
 \right\rbrace \, .
\end{multline}
In this expression, $g$ is the Yang-Mills coupling constant. 
The field content of the theory is the following.  
There are three complex scalars $\Phi_a = \phi_{2a-1} + i \phi_{2a}$ with $a \in \lbrace 1,2,3 \rbrace ,$ built from the real scalars transforming in the $\mathbf{6}$ representation of the R-symmetry group $ \mathrm{SO}(6)\simeq \SU(4).$   
We have four Weyl fermions $\psi^A$ with $A \in \lbrace 1,2,3,4 \rbrace$ transforming in the representation $\mathbf{4}$ of  $\SU(4) .$
Finally, the field strength is defined as
\beq
F_{\mu\nu} = \partial_{\mu} A_{\nu} - \partial_{\nu} A_{\mu} + i g [A_{\mu} , A_{\nu}] \, ,
\eeq
and the corresponding covariant derivatives $D_{\mu}$ read
\bea
& D_{\mu} \Phi_a = \partial_{\mu} \Phi_a + i g [A_{\mu}, \Phi_a] \, , & \\
& D_{\mu} \psi^A = \nabla_{\mu} \psi^A + i g [A_{\mu}, \psi^A] \, , &
\eea
where $\nabla_{\mu}$ is the covariant derivative on the three-sphere, {\sl i.e.} it contains the spin connection contribution when acting on the fermions.
The $C^{a}_{AB}$ are Clebsch-Gordan coefficients coupling two $\mathbf{4}$  representations and one $\mathbf{6}$ representation of the R-symmetry group $\SU(4).$
All the fields in the action transform under the adjoint representation of the gauge group $\SU(N)$.
The action is canonically normalized on the $\mathbb{R} \times S^3$ background with the radius of the three-sphere set to unity.

The classical Hamiltonian is obtained by performing the Legendre transform of the action \eqref{eq:N4SYMaction}.
We then decompose the fields into spherical harmonics on the three-sphere \cite{Ishiki:2006rt}, according to the conventions summarized in Appendix \ref{app:sphere_red}.
In this expansion a crucial role is played by the gauge field, since part of its degrees of freedom decouple on-shell in the near-BPS limit \eqref{eq:SMT_limits}, and they mediate an effective interaction at order $g^2$ in the coupling constant. 
The details of the decoupling of the gauge field and the corresponding Dirac quantization are extensively reported in Section 2.1 of \cite{Baiguera:2020jgy}, and similar discussions were presented in \cite{Baiguera:2020mgk,Baiguera:2021hky}.
Here we briefly highlights the main steps of the procedure.

The unphysical degrees of freedom of the gauge field are captured by the temporal and longitudinal components of the gauge field; they can be integrated out by using the Coulomb gauge $\nabla_i A^i =0$.
In order to keep track of the constraints, we consider a generic quadratic action in the field strength with the inclusion of a source
\begin{equation}
S_A = \int_{\mathbb{R}\times S^3} \sqrt{-\mathrm{det} \, g_{\mu\nu}} \, \tr(-\frac{1}{4} F_{\mu\nu}^2 - A^\mu j_\mu)
\, .
\end{equation}
After expanding the fields into spherical harmonics, the constraints become algebraic and we can express the result only in terms of the physical degrees of freedom, yielding the unconstrained Hamiltonian
\begin{equation}
\label{eq:Ham_freeYM}
H_A = \tr\sum_{J,m,\tilde{m}} \left[ \sum_{\rho = \pm1}\left( \frac{1}{2} |\Pi_{(\rho)}^{Jm\tilde{m}}|^2 + \frac{1}{2} \omega_{A,J}^2 |A_{(\rho)}^{Jm\tilde{m}}|^2+ A_{(\rho)}^{Jm\tilde{m}} j_{(\rho)}^{\dagger\,Jm\tilde{m}}\right) + \frac{1}{8J(J+1)} |j_0^{Jm\tilde{m}}|^2 \right] \,.
\end{equation}
The currents entering this expression can be identified by looking at the $\mathcal{N}=4$ SYM action reported in eq.~\eqref{eq:app_full_interacting_N=4SYM_Hamiltonian}.
One can now restore the interactions (including the other matter fields) to obtain the full Hamiltonian.
In order to proceed, we follow these steps:
\begin{enumerate}
	\item Determine the propagating modes in the near-BPS limit from the quadratic classical Hamiltonian.
	\item Derive the form of the currents that couple to the gauge fields by inspection of the $\mathcal{N}=4$ SYM action in eq.~\eqref{eq:app_full_interacting_N=4SYM_Hamiltonian}.
	\item Integrate out additional non-dynamical modes that give rise to effective interactions in the near-BPS limit.
	\item Derive the interacting Hamiltonian by taking the near-BPS limit in eq.~\eqref{eq:full_ham_gen} below.
\end{enumerate}
From now on, we focus specifically on the near-BPS limit \eqref{eq:SMT_limits} with $a_1=a_2=b_1=b_2=b_3=1,$ which characterizes the PSU(1,2|3) sector.
The decoupling limit can be written as
\begin{equation}
\label{eq:newnearBPSlimit}
g \rightarrow 0 \quad \mbox{with} \quad \frac{H - J}{g^2} \quad \mbox{fixed} \, , \qquad
J \equiv \mathbf{S}_1 + \mathbf{S}_2 + \sum_{i=1}^3 \mathbf{Q}_i \, ,
\end{equation}
with $N$ being fixed while sending $g \rightarrow 0$.
The interacting Hamiltonian describing the residual degrees of freedom of the sector is defined as
\begin{equation}
\label{eq:full_ham_gen}
 H_{\rm int} = \lim_{g \rightarrow 0} \frac{H - \mathbf{S}_1 - \mathbf{S}_2 - \sum_{i=1}^3  \mathbf{Q}_i}{g^2 N} \,.
\end{equation}
Let us focus on the contributions to the Hamiltonian mediated by the non-dynamical modes of the gauge field.
There is no contribution from the R-charges because the gauge field is neutral under such symmetry.
Therefore at quadratic order in the fields, the near-BPS limit involves the combination  
\begin{equation}
H_0 - \mathbf{S}_1 - \mathbf{S}_2 = 
\sum_{J,M}  \sum_{\rho=-1,1} \frac{1}{2} \left(|\Pi_{(\rho)}^{JM} - 2i\tilde{m}A_{(\rho)}^{\dagger\,JM}|^2+ (\omega_{A,J}^2 - 4 \tilde{m}^2) |A_{(\rho)}^{JM}|^2\right)
\, ,
\label{eq:quadratic_gauge_Hamiltonian_su12sectors}
\end{equation}
which must vanish. 
This corresponds to the constraint
\beq
\Pi^{JM}_{(\rho)} - 2 i \tilde{m} A^{\dagger \, JM}_{(\rho)} = 0 
\eeq
for all the non-dynamical modes of the gauge field, which means that we don't have to consider at the same time $\rho=-1$ and $|\tilde{m}|=J+1.$
After adding the sources written in eq.~\eqref{eq:Ham_freeYM} to the quadratic combination \eqref{eq:quadratic_gauge_Hamiltonian_su12sectors}, we find that the consistency of the constraints with the time evolution implies
\beq
\lbrace H, \Pi^{JM}_{(\rho)} - 2 i \tilde{m} A^{\dagger \, JM}_{(\rho)} \rbrace =
(\omega_{A,J}^2 - 4 \tilde{m}^2) A_{(\rho)}^{\dagger\,JM} + j_{(\rho)}^{\dagger\,JM} 
= 0 \, ,
\eeq
or equivalently
\beq
A_{(\rho)}^{Jm\tilde{m}} = -\frac{j^{ Jm\tilde{m}}_{(\rho)} }{\omega_{A,J}^2  -4\tilde{m}^2} \, .
\eeq
After using this additional constraint in eq.~\eqref{eq:quadratic_gauge_Hamiltonian_su12sectors} plus sources,
we find
\begin{equation}
H - \mathbf{S}_1 - \mathbf{S}_2 = \tr \left( \sum_{J,m,\tilde{m}} \frac{1}{8J(J+1)} |j_0^{Jm\tilde{m}}|^2 - \sum_{\rho=\pm1} \sum_{J,m,\tilde{m}}\frac{1}{2(\omega_{A,J}^2 - 4\tilde{m}^2)} |j_{(\rho)}^{Jm\tilde{m}}|^2\right)  \, .
\label{eq:general_gauge_mediated_interaction_su12limits}
\end{equation}
This expression will be used to compute all the gauge-mediated interactions, once the currents $j_0^{JM}, j^{JM}_{(\rho)}$ are identified from the interacting Hamiltonian of $\mathcal{N}=4$ SYM.

\subsection{Free Hamiltonian and reduction of the degrees of freedom}
\label{ssec:free_Ham_sphere_red}

As a first step in the spherical expansion method, we identify the degrees of freedom surviving the near-BPS limit by considering the combination $H_0 - J$ defined in eq.~\eqref{eq:newnearBPSlimit} at quadratic order in the fields.
A direct computation gives the following contributions for scalars $\Phi,$ gauge fields $A$ and fermions $\psi$:
\begin{align}
	\begin{split}
 \le H_0 - J \ri_{\Phi} &= \sum_{J,M} \sum_{a=1}^3 \tr \left( \left| (\Pi_a)_{JM} - i (2\tilde{m}+1) (\Phi_a^{\dagger})_{JM} \right|^2 + \left( \omega_J^2 - (2\tilde{m}+1)^2\right) (\Phi^\dagger_a)_{JM} (\Phi_a)_{JM} \right) \, ,  \\
 \le H_0 - J \ri_{A} &= \frac{1}{2}  \sum_{J,m,\tilde{m}} \sum_{\rho=-1,1} \tr \left(|\Pi_{(\rho)}^{JM} - 2i\tilde{m}A_{(\rho)}^{\dagger\,JM}|^2+ (\omega_{A,J}^2 - (2\tilde{m})^2) |A_{(\rho)}^{JM}|^2\right) \, ,  \\
\label{eq:quadratic_Ham_gauge field}
 \le H_0 - J \ri_{\psi} &= 
 \sum_{JM} \sum_{\kappa=\pm1} \tr \le  \le \kappa \omega^{\psi}_J + 2 \tilde{m} - \frac{3}{2} \ri (\psd_1)_{JM\kappa} (\psi_1)_{JM\kappa} \right.  \\
& \left.  - \sum_{A=2,3,4} \le \kappa \omega^{\psi}_J + 2 \tilde{m} + \frac{1}{2} \ri (\psd_A)_{JM\kappa} (\psi_A)_{JM\kappa} \ri \, . 
\end{split}
\end{align}
In the previous expressions, $\Pi_a$ represent the canonical momenta conjugate to the scalar fields, while $\Pi_{(\rho)}$ is the momentum associated to the gauge field $A_{(\rho)}.$
We present the consequences of imposing the condition $H_0-J=0$:
\begin{itemize}
\item \textbf{Scalars.} Given the definition $\omega_J = 2J+1$ and the constraint  $|\tilde{m}|\leq J,$ we find that all the three complex scalars with $a\in \lbrace 1,2,3 \rbrace$ have a surviving mode subject to the condition
\beq
(\Pi_a)_{J,m,J} + i \omega_J (\Phi^{\dagger}_a)_{J,m,J} = \mathcal{O}(g) \, .
\eeq
Furthermore, we remind that the restrictions on the angular momentum imply that $|m| \leq J.$
\item \textbf{Gauge fields.}
The dynamical gauge fields are the same of the $\SU(1,2)$ subsector, which was studied in \cite{Baiguera:2020mgk}. 
From the quadratic expression \eqref{eq:quadratic_Ham_gauge field} one deduces that all the components of the gauge field are non-dynamical, except when $\rho=-1$ and $\tilde{m}= \pm (J+1),$ with the constraints
\begin{equation}
\Pi_{(\rho)}^{Jm, \pm (J+1)} \pm i\omega_{A,J}A_{(\rho)}^{\dagger\,Jm,\pm (J+1)}=0  \, .
\label{eq:constraint_scalars}
\end{equation}
The modes with positive and negative eigenvalue for $\tilde{m}$ are related by the reality condition on the gauge field
\begin{equation}
  A_{(\rho=-1)}^{J,m,-J-1} = (-1)^{J-m} A_{(\rho=-1)}^{\dagger\,J,-m,J+1}  \,, 
 \label{eq:constraint_gauge}
\end{equation}
which allows to eliminate $A_{(\rho=-1)}^{J,m,-J-1}$ from all the expressions.
The surviving modes of the gauge field satisfy the momentum constraint $|m|\leq J.$
\item \textbf{Fermions.}
There are fermions of both chiralities $\kappa$ surviving the near-BPS limit \eqref{eq:newnearBPSlimit}.
When $\kappa=1,$ there is a surviving fermion field with R-symmetry index $A=1$ and momenta $|m|\leq J+ \frac{1}{2}$ together with fixed $\tilde{m}= -J .$
Therefore this particular field is
\beq
\psi^1_{J,m,-J, \kappa=1} \, . 
\label{eq:dynamical_modes_singlet_fermion_su123}
\eeq
There are more solutions for fermions with the other chirality $\kappa=-1,$ indeed we have the choices $A\in \lbrace 2,3,4 \rbrace$ for the $\SU(4)$ index.
They satisfy $|m|\leq J$ and have fixed $ \tilde{m} = J + \frac{1}{2} . $
We will collectively denote them as
\beq
 \psi^{A=2,3,4}_{J,m,J+\frac{1}{2}, \kappa=-1} \, . 
\label{eq:dynamical_modes_fermion_su123}
\eeq
All the fermions of $\mathcal{N}=4$ SYM survive, each of them with a particular value of the momentum $\tilde{m}.$
\end{itemize}
We observe that the dynamical modes of the scalar and field strength in eqs.~\eqref{eq:constraint_scalars} and \eqref{eq:constraint_gauge} satisfy an identity relating the conjugate momentum with the hermitian conjugate of a field.
This aspect is a non-relativistic trait typical of theories with Bargmann symmetry, which includes a $\mathrm{U}(1)$ central extension corresponding to the mass conservation.

The identities \eqref{eq:constraint_scalars} and \eqref{eq:constraint_gauge} constraining the modes of the bosonic fields are also responsible for the generation of non-trivial Dirac brackets, which read
\begin{align}
\{ (\Phi_a)_{J,m,J} , (\Phi^\dagger_a)_{J',m',J'} \}_D &=  \frac{i}{2\omega_J} \delta_{J,J'} \delta_{m,m'}  \, , 
\label{eq:Dirac_bracket_scalar_su122} \\
 \{(A_{\rho=-1})^{J,m,J+1}, (A_{\rho=-1}^{\dagger})^{J',m',J'+1}\}_D &= \frac{i}{2\omega_{A,J}} \delta_{J,J'} \delta_{m,m'} 
\, . 
\label{eq:Dirac_bracket_gauge_su122} 
\end{align}
In order to make the brackets canonically normalized, we redefine the dynamical bosonic modes surviving the near-BPS limit \eqref{eq:newnearBPSlimit} as
\beq
\Phi_{J,m}^a \equiv \sqrt{2 \omega_J} \Phi^a_{J,m,J} \, , \qquad
 A_{Jm} \equiv \sqrt{2\omega_{A,J}} A_{(\rho=1)}^{J,m,J+1}  \, .
\label{eq:dynamical_modes_bosons_su123}
\eeq
The fermions already have canonical Dirac brackets; we collect them using the following notation
\beq
 \psi^{A=1,2,3,4}_{Jm} = (\chi_{Jm}, \zeta^{a=1,2,3}_{Jm}) \, ,  
 \label{eq:collective_fermion_sphere_red}
\eeq
More explicitly, 
\beq
\begin{aligned}
& \chi_{Jm} \equiv \psi^1_{J,m,-J, \kappa=1} \, , \qquad
 \zeta^{a=1}_{Jm} \equiv \psi^{A=4}_{J,m,J+\frac{1}{2}, \kappa=-1} \, , & \\
& \zeta^{a=2}_{Jm} \equiv - \psi^{A=3}_{J,m,J+\frac{1}{2}, \kappa=-1} \, , \qquad
 \zeta^{a=3}_{Jm} \equiv \psi^{A=2}_{J,m,J+\frac{1}{2}, \kappa=-1}  \, . &
 \end{aligned}
 \label{eq:dynamical_fermions_psu123}
\eeq
The decomposition \eqref{eq:collective_fermion_sphere_red} clearly distinguishes between the fermions $\chi$ and $\zeta^a$, which transform under the global $\SU(3)$ residual R-symmetry either as a singlet or a triplet, respectively.

The fields \eqref{eq:dynamical_modes_bosons_su123} and \eqref{eq:collective_fermion_sphere_red} represent the full set of dynamical degrees of freedom surviving the PSU(1,2|3) near-BPS limit.
We identify them with the letters classified in \cite{Harmark:2007px}, where the general structure of all SMTs was considered.
The matching can be easily performed for scalars and gauge fields, giving the letters
\beq
A_{J,m} \rightarrow |d_1^n d_2^k \bar{F}_+ \rangle \, , \qquad
\Phi^1_{J,m} \rightarrow |d_1^n d_2^k Z \rangle \, , \qquad
\Phi^2_{J,m} \rightarrow |d_1^n d_2^k X \rangle \, , \qquad
\Phi^3_{J,m} \rightarrow |d_1^n d_2^k W \rangle \, .
\eeq
The matching for the fermions is less trivial.
The $\SU(4)$ R-symmetry charges easily allow to identify
the fermionic modes in the triplet with the letters of $\mathcal{N}=4$ SYM surviving the limit: they are
\beq
\zeta^1_{J,m} \rightarrow |d_1^n d_2^k \bar{\chi}_3 \rangle \, , \qquad
\zeta^2_{J,m} \rightarrow |d_1^n d_2^k \bar{\chi}_5 \rangle \, , \qquad
\zeta^3_{J,m} \rightarrow |d_1^n d_2^k \bar{\chi}_7 \rangle \, .
\eeq
The field $\chi_{Jm}$ secretly encodes two letters: this can be seen explicitly when considering the limit $J=0,$ which corresponds to the restriction to the $\SU(2|3)$ sector \cite{Baiguera:2021hky}.
In such case, the fermionic field has R-charges $(\frac{1}{2}, \frac{1}{2}, \frac{1}{2}),$ but there are two possible eigenvalues for the rotation operators, corresponding to $m = \pm \frac{1}{2} .$ 
The two cases are
\beq
\chi_{J=0,m=\frac{1}{2}} \rightarrow |\chi_2 \rangle \, , \qquad
\chi_{J=0,m=- \frac{1}{2}} \rightarrow | \chi_1 \rangle \, .
\eeq
From this observation, we understand that when the covariant derivatives are added, the field $\chi_{Jm}$ encodes information about two fermionic letters:
\beq
\chi_{Jm} \rightarrow 
|d_1^n d_2^k \chi_1  \rangle \, , 
|d_1^n d_2^k \chi_2 \rangle \, .
\eeq
These correspond to the non-trivial descendants related to the ancestor fermion mentioned in eq.~\eqref{eq:ancestor_fermion}.

Now we come back to the evaluation of the effective Hamiltonian of the system.
On the constraint surface, we find
\begin{equation}
\begin{aligned}
H_0  = \sum_{s=0}^{\infty}  \sum_{m=-\frac{s}{2}}^{\frac{s}{2}} & \left[ \sum_{a=1}^3  \le s+1 \ri \tr |\Phi^a_{s,m}|^2 
+ \le s + \frac{3}{2} \ri |\chi_{s,m}|^2 \right. \\
& \left. + \sum_{a=1}^3  \le s + \frac{3}{2} \ri |\zeta^a_{s,m}|^2
+ (s+2) \tr |A_{s,m}|^2 \right]
\, .
\end{aligned}
\label{H0_su123}
\end{equation}

\subsection{Purely bosonic interactions}
\label{ssec:bosonic_interactions_sphere_red}

We apply the general procedure outlined in Section \ref{ssec:general_procedure_sphere_red} to compute the interacting Hamiltonian according to eq.~\eqref{eq:full_ham_gen}.
The landscape of possible interactions can be splitted into three main categories:
\begin{itemize}
\item Terms mediated by the non-dynamical modes of the gauge field via the currents contributing to eq.~\eqref{eq:general_gauge_mediated_interaction_su12limits}.
It turns out that this class of terms nicely combines with quartic interactions where all the modes are bosonic and dynamical.
\item Cubic Yukawa terms.
\item Terms mediated by non-dynamical scalars or fermions and containing at least one dynamical gauge field. 
\item Quartic terms in the $\mathcal{N}=4$ SYM Hamiltonian containing only dynamical modes.
\end{itemize}

\noindent
The calculations for the spherical expansion will involve the summation of Clebsch-Gordan coefficients on the three-sphere.
In particular, the dynamical modes listed in eqs.~\eqref{eq:dynamical_modes_bosons_su123}, \eqref{eq:collective_fermion_sphere_red} and \eqref{eq:dynamical_fermions_psu123} fix in several computations the corresponding momenta of the Clebsch-Gordan coefficients.
For this reason, we introduce the following compact notation
\bea
\label{eq:notation_index_CJ_CM}
& (\mathcal{J},\mathcal{M}) \equiv (J, m, J) \, , \qquad
|m| \leq J  \, , &  \\
\label{eq:notation_index_hatCJ_hatCM}
& (\hat{\mathcal{J}}, \hat{\mathcal{M}}) \equiv \le J, m,J+1, \rho=-1 \ri \, , \qquad
|m| \leq J \, ,  & \\
\label{eq:notation_index_barCJ_barCM}
& (\bar{\mathcal{J}}, \bar{\mathcal{M}}) \equiv \le J, m, J+ \frac{1}{2}, \kappa=-1 \ri \, , \qquad
|m| \leq J \, .  &  
\eea
These saturations correspond to scalars, gauge fields and fermions, respectively.
We also introduce the short-hand notations
\begin{eqnarray}
&& \Delta J \equiv J_2 - J_1 = J_3 - J_4 \, , \qquad
\Delta m \equiv  m_2 - m_1 = m_3 - m_4 \, ,
\label{eq:def_deltaJ_deltam}
\\
&& s_i \equiv 2 J_i \, , \qquad
l \equiv 2 \Delta J \, , \qquad
|m_i| \leq \frac{s_i}{2} \, , \qquad
|\Delta m| \leq \frac{l}{2} \, ,
\label{eq:def_s_l}
\end{eqnarray}
which refer to an assignment of momenta that will often enter the results of the summation over Clebsch-Gordan coefficients.

At this point, we are ready to start the systematic computation of the effective Hamiltonian using the spherical expansion.
The procedure was reported and explained in detail in Section 2 of \cite{Baiguera:2020jgy}, Section 3 of \cite{Baiguera:2020mgk} and Section 3 of \cite{Baiguera:2021hky} for several subsectors.
Many of the calculations for the general PSU(1,2|3) case are similar; the main difference is that more terms are now involved and some generalizations to include them are needed.
Since the main changes are only technical, in the remaining part of the Section we will rather focus on a relevant subset of the possible interactions.
More specifically, we will only consider purely bosonic interactions because we argued in Section \ref{ssec:uniqueness_ham} that they are sufficient to uniquely fix the full effective Hamiltonian of the near-BPS limit.
Then we will proceed in Section \ref{ssec:comparison_cubic_supercharge} with the comparison between the results of the spherical expansion and of the cubic supercharge method, derived in eq.~\eqref{eq:Hint_tot_cubic}, showing that the purely bosonic terms indeed match.

According to the distinction between terms listed in the set of bullets above, there are four classes.
However, it is clear that the cubic Yukawa terms only generate interactions by integrating out an auxiliary field.
Since the Yukawa term contains two fermions and one scalar, it is clear that it will never contribute to purely bosonic interactions. 
Therefore, in the following we will only deal with three classes of terms.

\subsubsection*{Terms mediated by non-dynamical gauge field}

The terms mediated by the non-dynamical modes of the gauge field all contribute via eq.~\eqref{eq:general_gauge_mediated_interaction_su12limits}.
The main ingredient is represented by the currents, whose full expressions read
\begin{align}
\label{eq:current_j0_su123}
\begin{split}
& j_0^{\dagger\,Jm\tilde{m}} = \sum_{J_i m_i} \sum_{a=1}^3 \frac{g( \omega_{J_1}+\omega_{J_2})}{2\sqrt{\omega_{J_1}\omega_{J_2}}}  {\cal C}^{\CJ_2 \CM_2}_{\CJ_1 \CM_1; Jm\tilde{m}}  [(\Phi_a)_{J_1 m_1},(\Phi^\dagger_a)_{J_2 m_2}]  \\
& +  g \sum_{J_i m_i}  \mathcal{F}^{\bar{\CJ}_1 \bar{\CM}_1}_{\bar{\CJ}_2 \bar{\CM}_2; JM}  
\lbrace \chi_{J_1 m_1} , \chi^{\dagger}_{J_2 m_2} \rbrace
+  g \sum_{J_i m_i} \sum_{a=1}^3  \mathcal{F}^{\bar{\CJ}_2 \bar{\CM}_2}_{\bar{\CJ}_1 \bar{\CM}_1; JM}  
\lbrace (\zeta^{\dagger}_a)_{J_1 m_1} , (\zeta^a)_{J_2 m_2} \rbrace \\ 
& -  \sum_{J_i m_i} \frac{g}{2} \frac{\omega_{A,J_1}+\omega_{A,J_2}}{\sqrt{\omega_{A,J_1}\omega_{A,J_2}}}{\cal D}^{\hat{\CJ}_2 \hat{\CM}_2}_{JM; \hat{\CJ}_1 \hat{\CM}_1} [A_{J_1 m_1},A^\dagger_{J_2 m_2}]\,, 
\\
& j_{(\rho)}^{\dagger\,Jm\tilde{m}} = 
- 2g \sum_{J_i m_i} \sum_{a=1}^3 \sqrt{\frac{J_1(J_1+1)}{\omega_{J_1}\omega_{J_2}}} {\cal D}^{\CJ_2 \CM_2}_{\CJ_1 \CM_1 ; Jm\tilde{m},\rho}[(\Phi_a)_{J_1 m_1}, (\Phi^\dagger_a)_{J_2 m_2}] \\ 
& + g  \sum_{J_i, m_i}   \mathcal{G}^{\bar{\CJ}_1 \bar{\CM}_1}_{\bar{\CJ}_2 \bar{\CM}_2; JM \rho} 
\lbrace \chi_{J_1 m_1} , \chi^{\dagger}_{J_2 m_2} \rbrace  \\
& - g  \sum_{J_i, m_i} \sum_{a=1}^3  \mathcal{G}^{\bar{\CJ}_2 \bar{\CM}_2}_{\bar{\CJ}_1 \bar{\CM}_1; JM, -\rho} 
\lbrace (\zeta^{\dagger}_a)_{J_1 m_1} , (\zeta^a)_{J_2 m_2} \rbrace  \\
 & +  \frac{ig}{2}  \sum_{J_i m_i} \frac{\rho \omega_{A,J} - \omega_{A,J_1} - \omega_{A,J_2}}{\sqrt{\omega_{A,J_1}\omega_{A,J_2}}}
 {\cal E}_{JM\rho;\hat{\CJ}_1 \hat{\CM}_1}{}^{\hat{\CJ}_2 \hat{\CM}_2} [A_{J_1 m_1},A^\dagger_{J_2 m_2}] \, .
 \end{split}
%\label{eq:current_jrho_su123}
\end{align}
The currents were already simplified by applying the constraints \eqref{eq:constraint_scalars} and \eqref{eq:constraint_gauge} on the dynamical bosonic modes, and we used the notation introduced in eqs.~\eqref{eq:notation_index_CJ_CM}--\eqref{eq:notation_index_barCJ_barCM}.
Details on the Clebsch-Gordan coefficients are collected in Appendix \ref{app-definition_Clebsch}.

The terms mediated by non-dynamical gauge fields only contain double trace operators under the residual $\SU(3)$ R-symmery.
In order to find a closed form for the summation over Clebsch-Gordan coefficients, we need to include here the contributions coming from other quartic terms in the $\mathcal{N}=4$ SYM action:
\begin{itemize}
\item The scalar quartic self-interaction.
This is given by
\beq
 \frac{g^2}{2} \tr \le \frac{1}{2} |[\Phi_a, \Phi^{\dagger}_a]|^2 + |[\Phi_a, \Phi_b]|^2 \ri \, .
 \label{eq:purely_dynamical_scalar_quartic}
\eeq
The contribution to the gauge current arises only from the singlet part; the other term, giving a single trace operator, will be considered later.
\item The mixed scalar/gauge quartic interaction, which reads
\beq
- g^2 \tr \le [\Phi^{\dagger}_a, A^i] [A_i, \Phi_a] \ri \, , 
\eeq
and after the expansion into spherical harmonics becomes
\beq
\begin{aligned}
& g^2 \sum_{J M} \sum_{J_i m_i \rho_i} (-1)^{m-\tilde{m}+ m_4 -\tilde{m}_4} \mathcal{C}^{JM}_{J_1 M_1;J_4, -M_4} \mathcal{D}^{J, -M}_{J_2 M_2 \rho_2; J_3 M_3 \rho_3} \\
& \times  \tr \le  [ (\Phi_a^{\dagger})^{J_4 M_4} , A_{J_3 M_3 \rho_3}] [A_{J_2 M_2 \rho_2}, (\Phi_a)^{J_1 M_1}] \ri \, .
\end{aligned}
\label{eq:purely_dynamical_gauge_quartic}
\eeq
Using the cyclicity properties of the trace, we extract a double trace contribution from this interaction.
\end{itemize} 
As already mentioned below eqs.~\eqref{eq:notation_index_CJ_CM}--\eqref{eq:notation_index_barCJ_barCM}, we will only focus on the purely bosonic interactions.
To this aim, it means that we can consistently set to zero the fermionic fields inside the currents \eqref{eq:current_j0_su123}. 
The procedure to compute a charge density contribution in the spherical expansion language was extensively applied and explained in Sections 3.2 and 3.3 of \cite{Baiguera:2020mgk}.
Indeed, eq.~(3.80) of such reference contains precisely a charge density expression plus additional terms which vanish due to Gauss' law.

The computation for the present case is completely analog;
for the sake of simplicity, here we only report the main result and  we refer the reader to reference \cite{Baiguera:2020mgk} for the full derivation.
The purely bosonic part of the interactions mediated by the non-dynamical gauge field reads
\beq
 \frac{1}{2N} \sum_{l=1}^{\infty} \sum_{\Delta m=-l/2}^{l/2} \frac{1}{l}
 \tr \le \mathbf{Q}_{l, \Delta m}^{\dagger} \mathbf{Q}_{l, \Delta m} \ri \, ,
\label{eq:charge_density_sphere_red}
\eeq
where we define
\beq
\begin{aligned}
 \mathbf{Q}_{l,\Delta m} & \equiv \sum_{s_1=0}^{\infty} \sum_{m_1=-\frac{s_1}{2}}^{\frac{s_1}{2}} 
\le  \sum_{a=1}^3 
  C^{\frac{s_1+l}{2}, m_1+\Delta m}_{\frac{s_1}{2}, m_2; \frac{l}{2},\Delta m} [(\Phi_a^{\dagger})_{s_1 m_1}, (\Phi^a)_{s_1+ l, m_1 + \Delta m}] \right. \\
  & \left.  + C^{\frac{s_1+l}{2}, m_1 + \Delta m}_{\frac{s_1}{2}, m_1; \frac{l}{2}, \Delta m}
\sqrt{\frac{(s_1+1)(s_1+2)}{(s_1+l+1)(s_1+l+2)}}
 [ A^{\dagger}_{s_1 m_1}, A_{s_1+l, m_1+ \Delta m} ] \ri
   \, .
  \end{aligned}
\eeq 
Here we used the conventions summarized in eqs.~\eqref{eq:def_deltaJ_deltam} and \eqref{eq:def_s_l}.
The result written in the form \eqref{eq:charge_density_sphere_red} already takes into account the application of Gauss' law, which sets to zero various terms proportional to the total $\SU(N)$ charge.

\subsubsection*{Terms mediated by non-dynamical scalars}

The next set of interactions arises from the following cubic interaction of the $\mathcal{N}=4$ SYM Hamiltonian
\beq
- 4g\sqrt{J_1(J_1+1)} {\cal D}^{J_2 M_2}_{J_1 M_1 0, JM\rho} \tr \le A^{JM}_{(\rho)}[(\Phi_a)_{J_1 M_1},(\Phi^\dagger_a)_{J_2 M_2}]  \ri \, .
\eeq
While in the previous subsection we integrated out auxiliary modes of the gauge field, now we use instead a non-dynamical scalar field to mediate an effective interaction.
The result will be a quartic expression containing two scalars and two vectors.
This cases generalizes a similar term discussed in the SU(1,2|2) subsector, see eqs.~(3.95)-(3.99) of reference \cite{Baiguera:2020mgk}.
The result is
\beq
\frac{1}{2N} \sum_{l=0}^{\infty} \sum_{\Delta m=-l/2}^{l/2} 
\tr \le (W_a^{\dagger})_{l, \Delta m}  (W_a)_{l, \Delta m}  \ri \, ,
\label{eq:W_contribution_sphere}
\eeq
where we introduce the block
\beq
(W_a)_{l, \Delta m} =  \sum_{s_1=0}^{\infty} \sum_{m_1=-\frac{s_1}{2}}^{\frac{s_1}{2}}
\sqrt{\frac{l+1}{(s_1+l+1)(s_1+l+2)}}
 C^{\frac{s_1+m_1}{2}, m_1 +\Delta m_1}_{\frac{s_1}{2},m_1; \frac{l}{2},\Delta m} [A^{\dagger}_{s_1 m_1}, \Phi^a_{s_1+l, m_1 + \Delta m}] \, .
\eeq
The notation for the momenta used to present these expressions refers to the definitions \eqref{eq:def_deltaJ_deltam} and \eqref{eq:def_s_l}.

\subsubsection*{Single trace quartic scalar interaction}

We consider the single trace contribution arising from the quartic scalar interaction \eqref{eq:purely_dynamical_scalar_quartic}.
Expanding the fields into spherical harmonics and using the property
\beq
(\mathcal{Y}_{J_1 M_1})^* (\Omega) (\mathcal{Y}_{J_2 M_2})^* (\Omega) = 
\sum_{J_1, M_1, J_2, M_2} \mathcal{C}^{J M}_{J_1 M_1; J_2 M_2} 
(\mathcal{Y}_{J M})^* (\Omega) \, ,
\eeq
we reduce the number of spherical harmonics from four to three.\footnote{We refer to Appendix \ref{app-definition_Clebsch} for more details on the spherical harmonics. }
At this point, one finds by direct computation that the single trace scalar interaction becomes 
\beq
\frac{1}{4N} \sum_{JM,J_i} \frac{{\cal C}^{J,m,\tilde{m}}_{\mathcal{J}_1 \mathcal{M}_1;\mathcal{J}_2 \mathcal{M}_2}{\cal C}^{J,m,\tilde{m}}_{\mathcal{J}_3 \mathcal{M}_3; \mathcal{J}_4 \mathcal{M}_4}}{\sqrt{ \omega_{J_1} \omega_{J_2} \omega_{J_3} \omega_{J_4}}}\tr([(\Phi_a)_{J_1 m_1}, (\Phi_b)_{J_2 m_2}][(\Phi_b^\dagger)_{J_3 m_3},(\Phi_a^\dagger)_{J_4 m_4}]) \, .
\eeq
It is easy to find that the constraints on momenta and the triangle inequality 
\beq
\tilde{m} = J_1 + J_2 = J_3 + J_4 \, ,
\qquad
\mathrm{Max} \lbrace |J_1 - J_2|, |J_4-J_3| \rbrace \leq J \leq J_1 + J_2 
\eeq
give the saturation condition 
\beq
J= J_1 + J_2 = J_3 + J_4 \, .
\eeq
Thus the sum over intermediate momenta collapses and we obtain a contribution only from this fixed value of $J.$
We get
\beq
 \begin{aligned}
H_{\rm int} & = \frac{1}{4N} \sum_{s_i=0}^{\infty}
\sum_{m_i=-\frac{s_i}{2}}^{\frac{s_i}{2}}
\frac{\delta^{s_1+s_2}_{s_3+s_4}}{s_1+s_2+1}  \,
 C^{s_1+s_2, m_1+m_2}_{s_1 m_1; s_2 m_2} 
 C^{s_3+s_4, m_3+m_4}_{s_3 m_3; s_4 m_4} 
\\
& \times  \tr([(\Phi_a)_{s_1 m_1},(\Phi_b)_{s_2 m_2}][(\Phi_b^\dagger)_{s_3 m_3},(\Phi_a^\dagger)_{s_4 m_4}])\,.
\end{aligned} 
\label{eq:single_trace_term_scalars_sphere}
\eeq

\subsection{Comparison with the cubic supercharge method}
\label{ssec:comparison_cubic_supercharge}

Summing the contributions \eqref{eq:charge_density_sphere_red}, \eqref{eq:W_contribution_sphere} and \eqref{eq:single_trace_term_scalars_sphere} from bosonic modes to the interacting Hamiltonian, we finally obtain
\beq
\begin{aligned}
H_{\rm int,bos} & =  \frac{1}{2N}
 \sum_{l=1}^{\infty}  \sum_{\Delta m=-\frac{l}{2}}^{\frac{l}{2}} \tr \le \mathbf{Q}_{l, \Delta m}^{\dagger} \mathbf{Q}_{l, \Delta m} \ri + \sum_{l=0}^{\infty}  \sum_{\Delta m=-\frac{l}{2}}^{\frac{l}{2}}
\tr \le (W_a^{\dagger})_{l, \Delta m}  (W_a)_{l, \Delta m}  \ri  \\
& + \frac{1}{4N} \sum_{s_i=0}^{\infty}
\sum_{m_i=-\frac{s_i}{2}}^{\frac{s_i}{2}}
\frac{\delta^{s_1+s_2}_{s_3+s_4}}{s_1+s_2+1}  \,
 C^{s_1+s_2, m_1+m_2}_{s_1 m_1; s_2 m_2} 
 C^{s_3+s_4, m_3+m_4}_{s_3 m_3; s_4 m_4} \\
& \times   \tr([(\Phi_a)_{s_1 m_1},(\Phi_b)_{s_2 m_2}][(\Phi_b^\dagger)_{s_3 m_3},(\Phi_a^\dagger)_{s_4 m_4}])\,.
\end{aligned}
\label{eq:ham_bos_sphere}
\eeq
We compare this expression with the bosonic contributions included in the Hamiltonian \eqref{eq:Hint_tot_cubic} obtained using the cubic supercharge method.
The dictionary to match the momenta $(J_i, m_i)$ used in the spherical expansion computation with the integer numbers $(n,k)$ labelling the $\SU(1,2)$ representation is given by
\bea
& n = \Delta J - \Delta m \, , \qquad
k = \Delta J + \Delta m \, , \qquad
n'=J_1 - m_1 \, , & \\
& k'=J_1 + m_1 \, , \qquad
n+n'= J_2 - m_2 \, , \qquad
k+k'= J_2 + m_2 \, . &
\eea
Applying these relations, one can check that the blocks $\mathbf{Q}_{nk}$ and $(W_a)_{nk}$ (defined in eqs.~\eqref{eq:charge_density_sphere_red} and \eqref{eq:W_contribution_sphere} via the spherical expansion) coincide with the homonymous blocks in eqs.~\eqref{eq:def_charge_density_cubicgen} and \eqref{eq:definition_FKHWM} (defined via the cubic supercharge method), respectively.

Finally, one can also show that the scalar F-term obtained from the spherical expansion technique coincides with the result from the cubic supercharge.
To this aim, it is sufficient to notice that
\beq
 \delta^{n+n'}_{p+p'} \delta^{k+k'}_{q+q'} \,
P^{(1,2)}_{n,k;n',k'}
P^{(1,2)}_{p,q;p',q'}  =
\frac{\delta^{s_1+s_2}_{s_3+s_4}}{s_1+s_2+1} \,
  C^{s_1+s_2,m_1+m_2}_{s_2 m_2; s_1 m_1} \, 
  C^{s_3+s_4,m_3+m_4}_{s_3 m_3; s_4 m_4} \, .
\eeq
The two sides of the equality correspond to the non-trivial coefficients of the scalar F-term as computed from the two methods.
The left-hand side comes from the contribution of the block defined in eq.~\eqref{eq:block_F2}, to the Hamiltonian \eqref{eq:Hint_tot_cubic}.
The right-hand side is the coefficient of the last term in eq.~\eqref{eq:ham_bos_sphere}.
The apparent different structure of the fields under the residual $\SU(3)$ R-symmetry also matches after using the property $\epsilon^{eab} \epsilon^{ecd} = \delta^{ac} \delta^{bd} - \delta^{ad} \delta^{bc} $ of the Levi-Civita symbol, and the antisymmetry of the commutators.

This concludes the matching between the bosonic part of the spherical expansion and of the cubic supercharge methods, thus providing a non-trivial check of the computations.
Furthermore, following the argument presented in Section \ref{ssec:uniqueness_ham}, this result is sufficient to guarantee the uniqueness of the interacting Hamiltonian derived in eq.~\eqref{eq:Hint_tot_cubic}.

\section{Discussion}
\label{sec:discussion}

In this paper, we constructed a supercharge which is cubic in terms of the letters in the PSU$(1,2|3)$ sector of $\mN=4$ SYM. 
This is the generalization of the cubic supercharge in the PSU$(1,1|2)$ subsector \cite{Zwiebel:2007cpa,Beisert:2007sk}.
We use the supercharge \eqref{eq:linear_combination_fermgen} to derive a manifestly positive definite Hamiltonian invariant under the action of all the generators of PSU$(1,2|3)$: 
\begin{align}\label{eq:Hint_tot_cubic-DISCUSSION}
	\begin{split}
		H_{\rm int} &= H_D + H_F \, ,
		\\
		H_D &=	\sum_{n,k=0}^{\infty}
		\tr \left[ (\mathcal{B}_0^{\dagger})_{n,k} (\mathcal{B}_0)_{n,k}  + \sum_{a=1}^3 \sum_{I=1,2} (\mathcal{B}_I^{a\dagger})_{n,k} (\mathcal{B}_I^a)_{n,k}  + (\mathcal{B}_3^{\dagger})_{n,k} (\mathcal{B}_3)_{n,k} \right] \, ,  \\
		H_F	&= \sum_{n,k=0}^{\infty}
		\tr \left[ (\mathcal{F}_0^{\dagger})_{n,k} (\mathcal{F}_0)_{n,k}  + \sum_{a=1}^3 \sum_{I=1,2} (\mathcal{F}_I^{a\dagger})_{n,k} (\mathcal{F}_I^a)_{n,k}  + (\mathcal{F}_3^{\dagger})_{n,k} (\mathcal{F}_3)_{n,k} \right]   \, ,
	\end{split}
\end{align}
This positive definite form \eqref{eq:Hint_tot_cubic-DISCUSSION} shows the Hamiltonian is composed by fundamental blocks including both D-term blocks $\mathcal{B},$ defined in eqs.~\eqref{eq:block_B0}--\eqref{eq:block_B3}, and F-term blocks $\mathcal{F}$ introduced in eqs.~\eqref{eq:block_F0}--\eqref{eq:block_F3}. 
Each kind of block is transforming as a supermultiplet, which can be considered as a manifestation of the supersymmetry invariance. 
Both the $H_D$ and $H_F$ terms are invariant under supersymmetry separately, indicating an enhancement of supersymmetry. 
This construction resolves our puzzles about Spin Matrix theory in the PSU$(1,1|2)$ subsector \cite{Baiguera:2021hky}, where one lacks the knowledge of how to reorganize the Hamiltonian into a manifest positive definite form.

A powerful tool to investigate supersymmetric field theories consists in the construction of superspace, which is an extension of the standard spacetime with Grassmann coordinates to account for the graded structure of the super-algebra.
The advantages of this method are that all the standard field content of the system is encoded in the superfield, actions can be written in a simpler way, and SUSY is manifest at each step.
It is interesting to notice that the construction of a cubic supercharge for PSU(1,2|3) SMT of this paper is related to a cubic supercharge construction presented in \cite{Chang:2013fba,Chang:2022mjp}. 
This can be seen by introducing the fermionic superfield on the flat superspace $\mathbb{C}^{2|3}$ given by
\begin{eqnarray}
\Psi (\mathcal{Z}) &=& \sum_{n,k=0}^{\infty} z_+^n z_-^k \le \sqrt{\frac{(n+k-1)!}{n!k!}} \chi_{n,k} + \theta_a  \sqrt{\frac{(n+k)!}{n!k!}} (\Phi_a)_{n,k} \right.
\nn \\ && \left.
+ 2 \epsilon^{abc} \theta_a \theta_b  \sqrt{\frac{(n+k+1)!}{n!k!}}(\zeta_c)_{n,k}  +  \theta_1 \theta_2 \theta_3 \sqrt{\frac{(n+k+2)!}{n!k!}} A_{n,k}  \ri \, ,
\label{eq:superfield_us}
\end{eqnarray}
where we have defined the collective coordinate
\beq
\mathcal{Z} \equiv (z^i, \theta^a) \equiv (z^+, z^-, \theta_1, \theta_2, \theta_3) \, ,
\eeq
with $z^i$ being auxiliary bosonic coordinates and $\theta^a$ being the Grassmann coordinates defining the graded extension of spacetime.
The connection to the cubic supercharge  construction of \cite{Chang:2013fba,Chang:2022mjp} is now completed by noticing that the hermitian conjugate $\mathcal{Q}^{\dagger}$ of our cubic generator \eqref{eq:linear_combination_fermgen} obeys
\beq
\lbrace \mathcal{Q}^{\dagger}, \Psi(\mathcal{Z}) \rbrace_D = \Psi^2 (\mathcal{Z}) \, . 
\label{eq:cohomology_Qdag}
\eeq
This is in correspondence with the behavior of the superfield and cubic supercharge introduced in \cite{Chang:2013fba,Chang:2022mjp}. It would be interesting to explore this connection further, as it points to yet another way to find the effective Hamiltonian derived in eq.~\eqref{eq:Hint_tot_cubic}.
Moreover, the supercharge and superfield construction of \cite{Chang:2013fba,Chang:2022mjp} are closely related to the investigation of $1/16$--BPS operators of $\mathcal{N}=4$ SYM theory, as the 
space of $1/16$--BPS operators is isomorphic to the cohomology of such a supercharge \cite{Grant:2008sk}.

Superfields have been studied in the context of non-relativistic theories from a different perspective, to describe quantum field theories with either Lifshitz or Schroedinger invariance \cite{Meyer:2017zfg,Auzzi:2019kdd,Arav:2019tqm}.
The superfield formalism was recently applied to study the renormalization structure of three-dimensional SUSY Galilean Electrodynamics, which is an $\mathcal{N}=2$ supersymmetric gauge theory arising from the null reduction of $\mathcal{N}=1$ SYM in four dimensions \cite{Baiguera:2022cbp}.
One may hope to perform the null reduction of the full $\mathcal{N}=4$ SYM theory and compare with the SMT formalism.
This comparison is not only heuristic, but could be motivated by previous investigations which found a (semi)local formulation of SMTs with $\mathrm{SU}(1,1)$ symmetry subgroup as field theories living on a circle \cite{Baiguera:2020jgy}.
This included the introduction of a superspace in the case of sectors with supersymmetry invariance.
The counting of degrees of freedom and thermodynamic arguments suggest that the SMTs with $\mathrm{SU}(1,2)$ subgroup could be effectively described as 2+1 dimensional field theories, thus corresponding to the same number of dimensions of a null reduction of $\mathcal{N}=4$ SYM.
A more direct approach to find a (semi)local formulation of these SMTs could be based on the results presented in \cite{Lambert:2021nol}.

The Spin Matrix theory is also closely related to the chiral algebra of $\mN=4$ SYM, as the decoupling condition of PSU$(1,1|2)$ subsector is identical to the Schur condition \cite{Beem:2013sza,Bonetti:2016nma}.
The letters in the 2d chiral algebra descriptions are ghost-like fields and the central charge of the 2d chiral algebra is negative, as required by unitarity of 4d field theory. 
A similar phenomenon was also observed in the SU$(1,1|1)$ subsector of Spin Matrix theory \cite{Baiguera:2020jgy}, where the letters can be considered as excitations above the decoupling limit. 
Similar situation can also be observed in the PSU$(1,2|3)$ sector.
The blocks $W_I^\dagger$ and the letter $V_I$ are all transforming in the $(p,q)=(0,I-3)$ representations of SU$(1,2)$ algebra. 
However, there is an essential difference: the letters in $I=0,2$ representations are fermions, while the bosons transform in representations with $I =1,3$. 
This is completely opposite to the quantum statistical nature of the blocks. 
Therefore, studying Spin Matrix theory provides a generalization of chiral algebra theory and also novel insights to understand the nature of chiral algebra, from a theory with larger global symmetry.

The next ambitious project after the construction of Spin Matrix theory is to understand its gravity dual from various aspects.
It can be analysed from dual string theory in torsional Newton-Cartan geometry. 
Depending on the scaling dimension of the states, the dual gravitational description can either be in terms of giant gravitons \cite{Harmark:2016cjq}, a brane model \cite{Lin:2005nh} or a black hole-like geometry \cite{Kunduri:2006ek,Chong:2005da,Chong:2005hr}. 
Our work paves the way towards understanding the physics of $1/16$-BPS AdS black hole from the dual quantum mechanical theory. 
One of the most amazing achievement in recent years was the understandings of black hole entropy from the superconformal index computation \cite{Benini:2018ywd,Choi:2018hmj,Murthy:2020rbd,Goldstein:2020yvj}.
Our work provides an alternative way of understanding the microscopic states dual to the black holes, which can be acquired by solving the constraints 
\begin{equation}
	(\mathcal{B}_I)_{n,k} \ket{\Omega} = 	(\mathcal{F}_I)_{n,k} \ket{\Omega} =0
\end{equation}
Such study could potentially reveal the feature of black hole-like states in the dual field theory.
This is also visited in the recent work \cite{Chang:2022mjp}. 
An observation made in \cite{Choi:2018hmj} via the superconformal index computation is the possible existence of $1/8$-BPS black hole (which is called the PSU$(1,1|2)$ subsector in SMT) in the dual AdS gravity, although the analytic gravitational solutions were never found.
A common feature shared by PSU$(1,1|2)$ and PSU$(1,2|3)$ subsector is the simultaneous presence of both D-term blocks and the F-term blocks. 
It is unclear whether this fact is related to the black hole states but worthy to explore in the future.  
More importantly, the SMT interactions break the exact BPS conditions, which could potentially teach us the physics of near-BPS black holes \cite{Larsen:2019oll}. 

Our works \cite{Baiguera:2020jgy,Baiguera:2020mgk,Baiguera:2021hky} including this paper, have already developed various methods in constructing Hamiltonian of Spin Matrix theory \cite{Harmark:2014mpa}, as the low energy effective theory of $\mN=4$ SYM. 
One could test whether these methodologies are useful in studying the low energy limit of superconformal field theories in other dimensions. 
One example of interest is the $\mN=6$ superconformal Chern-Simons theory in $D=3$ \cite{Aharony:2008ug} (known as ABJM theory). 
It was discussed in \cite{Grignani:2008is} that the effective field theory in the SU$(2)\times$SU$(2)$ subsector are two decoupled Heisenberg spin chain models. 
Other larger subsectors are  discussed in \cite{Klose:2010ki}. 
Constructing the corresponding Spin Matrix theory in these sector could provide more examples of generalized solvable spin chain models.

Other interesting future applications of the PSU$(1,2|3)$ SMT include studying the coherent state \cite{Berenstein:2022srd,Holguin:2022drf,Lin:2022wdr}, the generalized magnetic Spin Matrix theory \cite{Harmark:2006ie}, computing the SMT partition functions and studying the possible modular properties of 4d partition functions \cite{Gadde:2020bov,Jejjala:2021hlt,Jejjala:2022lrm}.

\section*{Acknowledgements}

We thank Chi-Ming Chang and Sam van Leuven for useful discussions, and we thank the anonymous referee for insightful comments that improved the manuscript.
S.B. acknowledges support from the Israel Science Foundation (grant No.~1417/21), from the German Research Foundation through a German-Israeli Project Cooperation (DIP) grant ``Holography and the Swampland'', from  Carole and Marcus Weinstein through the BGU Presidential Faculty Recruitment Fund  and is grateful to the Azrieli foundation for the award of an Azrieli fellowship.
T.H. is supported by the Independent Research Fund Denmark grant number DFF-6108-00340 ``Towards a deeper understanding of black holes with non-relativistic holography''.
Y.L. was supported by the UCAS program of special research associate, the internal funds of the KITS, and the
Chinese Postdoctoral Science Foundation.
Y.L. is also supported by a Project Funded by the Priority Academic Program Development of Jiangsu Higher Education Institutions (PAPD).

\appendix

%%%%%%%%%%%%%%%%%
\section{Details on the PSU(1,2|3) invariance of the cubic generator}
\label{app:details_invariance_cubic}

We provide additional details about the invariance of the cubic supercharge \eqref{eq:linear_combination_fermgen} under the symmetry transformations of the spin group PSU(1,2|3).
The supercharges $Q_{a}$ defined in eq.~\eqref{eq:supercharge-asletters} act on the levels $n$ (which are the descendants of spatial direction $d_1$), while the supercharges of $\tilde{Q}_{a}$ act on the levels $k$  (which are the descendants of spatial direction $d_2$). 
Similarly, the supercharges of $S_{a}$ act on the levels $n+k$. 
These three different classes are precisely in relation with the three different momenta entering the saturated CG coefficients. 
We will list a few useful identities satisfied by $P^{(i,j)}_{n,k;n'k'}$ which can be easily checked by direct computation. 

\subsubsection*{Relations related to $Q^\dagger$ supercharges}
\begin{align}
	\begin{split}
& P^{(i,j)}_{n,k;n',k'} = P^{(j,i)}_{n',k';n,k}  \, ,  \qquad  \sqrt{n'} P^{(i,j+1)}_{n,k;n'-1,k'}+ \sqrt{n} P^{(i+1,j)}_{n-1,k;n'k'} = \sqrt{n+n'} P^{(i,j)}_{n,k;n',k'}   \\
& \sqrt{n+n'+1} P^{(i,j+1)}_{n,k;n',k'} = \sqrt{n'+1} P^{(i,j)}_{n,k;n'+1,k'}\,, \\
& \sqrt{n'+1} P^{(i+1,j)}_{n,k;n'+1,k'} = \sqrt{n+1} P^{(i,j+1)}_{n+1,k;n',k'}\,,  \quad  \sqrt{n+n'+1} P^{(i+1,j)}_{n,k;n',k'} = \sqrt{n+1} P^{(i,j)}_{n+1,k;n',k'} \\
& \sqrt{n-n'} P^{(i,j)}_{n',k';n-n',k-k'} = \sqrt{n} P^{(i,j+1)}_{n',k';n-1-n',k-k'} \\
& \sqrt{n'+1} P^{(i,j)}_{n'+1,k';n-1-n',k-k'}  =\sqrt{n} P^{(i+1,j)}_{n',k';n-1-n',k-k'} \\
& \sqrt{n'} P^{(i+1,j)}_{n'-1,k';n+1-n',k-k'} +\sqrt{n-n'+1} P^{(i,j+1)}_{n',k';n-n',k-k'} = \sqrt{n+1} P^{(i,j)}_{n',k';n+1-n',k-k'}
\label{eq:useful_identities_P-Qsupercharge}
	\end{split}
\end{align}

\subsubsection*{Relations related to $S$ supercharges}
\begin{align}\label{eq:useful_identities_P-Ssupercharge}
	\begin{split}
& \sqrt{n'+k'+j} P^{(i,j)}_{n;k;n',k'} = \sqrt{n+k+n'+k'+i+j} P^{(i,j+1)}_{n,k;n',k'} \\
& \sqrt{n+k+n'+k'+i+j} P^{(i,j)}_{n;k;n',k'} -\sqrt{n'+k'+j} P^{(i,j+1)}_{n,k;n',k'} =\sqrt{n+k+i}\, P^{(i+1,j)}_{n,k;n',k'}
	\end{split}
\end{align}

\subsection{Bosonic generators}
\label{app:bos_generators}

We start by showing that the generic structure \eqref{eq:generic_TA_term} with coefficients \eqref{eq:Pij_coefficients} is invariant under the action of the generators of the $\SU(1,2)$ subgroup.
As mentioned in Section \ref{ssec:construction_cubic_supercharge}, 
the assignment of labels $(n,k)$ and the representations $I$ under which the fields transform are already chosen in such a way to have a vanishing eigenvalue of the Cartan generators $L_0, \tilde{L}_0.$ 
Here we consider the action of the other non-diagonal generators, such as $L_+.$ We find
\begin{eqnarray}
\{ L_+ , T_A \}_D &=& \sum_{n,k,n',k'=0}^\infty P^{(i,j)}_{n,k,n',k'} \left[ \sqrt{(n+1)(n+k+i)} \tr ([V^\dagger_{n+1,k}  ,\tilde{V}^\dagger_{n',k'} \} \hat{V}_{n+n',k+k'}) \right.
\nn \\ &&
+\sqrt{(n'+1)(n'+k'+j)} \tr ([ V^\dagger_{n,k} , \tilde{V}^\dagger_{n'+1,k'} \} \hat{V}_{n+n',k+k'})
\nn \\ && \left.
- \sqrt{(n+n')(n+n'+k+k'+i+j-1)}\tr ([ V^\dagger_{n,k} , \tilde{V}^\dagger_{n',k'} \} \hat{V}_{n+n'-1,k+k'} )\right] =
\\ &=&\sum_{n,k,n',k'=0}^\infty  \left[ \sqrt{n(n+k+i-1)} P^{(i,j)}_{n-1,k,n',k'} + \sqrt{n'(n'+k'+j-1)} P^{(i,j)}_{n,k,n'-1,k'}  \right.
 \nn  \\ && \left.  - \sqrt{(n+n')(n+n'+k+k'+i+j-1)} P^{(i,j)}_{n,k,n',k'} \right]  \tr ([V^\dagger_{n,k}  ,\tilde{V}^\dagger_{n',k'} \}\hat{V}_{n+n'-1,k+k'}) = \nn \\
 & = & 0 \, . \nn
\end{eqnarray}
The terms in the first step correspond to the three possibilities to act with the $L_+$ generator on the cubic structure of fields in the structure \eqref{eq:generic_TA_term}. 
In the second step we shifted the labels of the first two terms as $n\rightarrow n-1$ and $n' \rightarrow n'-1,$ respectively, in order to collect the same trace structure.
Finally, the sum of the coefficients in the square parenthesis vanish after using the properties \eqref{eq:useful_identities_P-Qsupercharge}.
Using a similar procedure, one can explicitly check that the action of the other generators $L_-, J_+$ on the structure \eqref{eq:generic_TA_term} also vanishes.
Furthermore, since $T_A$ is symmetric in the indices $(n,k),$ the same steps also show that the bosonic generators $\tilde{L}_+, \tilde{L}_-, J_-$ commute with it.

The remaining bosonic generators of the sector belong to the $\SU(3)$ R-symmetry group.
However, it is easy to observe that the specific structures \eqref{T1}--\eqref{T6} are singlets, since they are built by using the invariant tensors $\delta^{ab}, \epsilon^{abc}$ of this group.

\subsection{Fermionic generators}
\label{app:ferm_generators}

Now we focus on the invariance under the supercharges $Q_{4-a}.$
As explained in Section \ref{ssec:construction_cubic_supercharge}, this result combined with the invariance under the $Q^{\dagger}_{4-a}$, derived in eq.~\eqref{eq:vanishing_SUSY_variation_fermgenQdag}, is sufficient to show the invariance under all the other fermionic generators.
Here we list the full set of Dirac brackets with the terms defined in eqs.~\eqref{T1}--\eqref{T6}:
\begin{equation}
\{ Q_{4-a},T_1 \}_D =  \sum_{n,k,n',k'=0}^\infty P^{(1,0)}_{n,k,n',k'} \sqrt{n+n'+1} \tr ( (\Phi^\dagger_a)_{n,k}\{ \chi^\dagger_{n',k'},\chi_{n+n'+1,k+k'} \} )  \, ,
\end{equation}
\begin{eqnarray}
\{ Q_{4-a},T_2 \}_D &=&  \sum_{n,k,n',k'=0}^\infty P^{(1,1)}_{n,k,n',k'} \sqrt{n+n'+1} \tr (  (\Phi^\dagger_a)_{n,k}[ (\Phi^\dagger_b)_{n',k'}, (\Phi_b)_{n+n'+1,k+k'} ] )
\nn \\ &&  -  \sum_{n,k,n',k'=0}^\infty P^{(0,2)}_{n,k,n',k'} \sqrt{n+n'+1}\, \epsilon^{abc} \tr ( \chi^\dagger_{n,k} [   (\zeta^\dagger_b)_{n',k'} , (\Phi_c)_{n+n'+1,k+k'}] )
\nn \\ &&
-\sum_{n,k,n',k'=0}^\infty P^{(1,0)}_{n,k,n',k'} \sqrt{n+n'+1} \tr ( (\Phi^\dagger_a)_{n,k}\{ \chi^\dagger_{n',k'},\chi_{n+n'+1,k+k'} \} ) \, ,
\end{eqnarray}
\begin{align}
	\begin{split}
\{ Q_{4-a},T_3 \}_D &=  \sum_{n,k,n',k'=0}^\infty P^{(1,2)}_{n,k,n',k'} \sqrt{n+n'+1} \tr (  (\Phi^\dagger_a)_{n,k}\{ (\zeta^\dagger_b)_{n',k'}, (\zeta_b)_{n+n'+1,k+k'} \} )
 \\ & -  \sum_{n,k,n',k'=0}^\infty P^{(0,3)}_{n,k,n',k'} \sqrt{n+n'+1} \tr ( \chi^\dagger_{n,k} [   A^\dagger_{n',k'} , (\zeta_a)_{n+n'+1,k+k'}  ] )
 \\ &
+\sum_{n,k,n',k'=0}^\infty P^{(0,2)}_{n,k,n',k'} \sqrt{n+n'+1} \epsilon^{abc} \tr ( \chi^\dagger_{n,k} [ (\zeta^\dagger_b)_{n',k'},(\Phi_c)_{n+n'+1,k+k'} \} )  \, ,
\\
\{ Q_{4-a},T_4 \}_D &=  \sum_{n,k,n',k'=0}^\infty P^{(1,3)}_{n,k,n',k'} \sqrt{n+n'+1} \tr (  (\Phi^\dagger_a)_{n,k} [ A^\dagger_{n',k'}, A_{n+n'+1,k+k'} ] )
 \\ & +  \sum_{n,k,n',k'=0}^\infty P^{(0,3)}_{n,k,n',k'} \sqrt{n+n'+1} \tr ( \chi^\dagger_{n,k} [   A^\dagger_{n',k'} , (\zeta_a)_{n+n'+1,k+k'}  ] ) \, ,
 \\
\{ Q_{4-a},T_5 \}_D &=  \sum_{n,k,n',k'=0}^\infty P^{(2,1)}_{n,k,n',k'} \sqrt{n+n'+1} \tr ( [ (\zeta^\dagger_b)_{n,k} ,(\Phi^\dagger_a)_{n',k'} ] (\zeta_b)_{n+n'+1,k+k'}  )
 \\ & -  \sum_{n,k,n',k'=0}^\infty P^{(2,1)}_{n,k,n',k'} \sqrt{n+n'+1} \tr ( [ (\zeta^\dagger_b)_{n,k} ,(\Phi^\dagger_b)_{n',k'} ] (\zeta_a)_{n+n'+1,k+k'}  )
 \\ & -  \sum_{n,k,n',k'=0}^\infty P^{(1,1)}_{n,k,n',k'} \sqrt{n+n'+1} \tr ( [ (\Phi^\dagger_a)_{n,k} ,  (\Phi^\dagger_b)_{n',k'} ] (\Phi_b)_{n+n'+1,k+k'} )  \, , 
\\
\{ Q_{4-a},T_6 \}_D &=  \sum_{n,k,n',k'=0}^\infty P^{(2,2)}_{n,k,n',k'} \sqrt{n+n'+1}\, \epsilon^{abc} \tr ( \lbrace (\zeta^\dagger_b)_{n,k} ,(\zeta^\dagger_c)_{n',k'} \rbrace A_{n+n'+1,k+k'}  )
 \\ & +  \sum_{n,k,n',k'=0}^\infty P^{(1,3)}_{n,k,n',k'} \sqrt{n+n'+1} \tr ( [ (\Phi^\dagger_a)_{n,k} ,A^\dagger_{n',k'} ] A_{n+n'+1,k+k'}  )
 \\ &  +  \sum_{n,k,n',k'=0}^\infty P^{(1,2)}_{n,k,n',k'} \sqrt{n+n'+1} \tr ( [ (\Phi^\dagger_b)_{n,k} ,(\zeta^\dagger_b)_{n',k'} ] (\zeta_a)_{n+n'+1,k+k'}  )   \, .
\end{split}
\end{align}
Using the cyclicity of the trace and the properties \eqref{eq:useful_identities_P-Qsupercharge}, most of the terms directly cancel when we build the linear combination \eqref{eq:linear_combination_fermgen}.
A non-trivial simplification comes from the following expression, which vanishes by antisymmetry:
\beq
\begin{aligned}
& \sum_{n,k,n',k'=0}^\infty P^{(2,2)}_{n,k,n',k'} \sqrt{n+n'+1}\, \epsilon^{abc} \tr ( \lbrace (\zeta^\dagger_b)_{n,k} ,(\zeta^\dagger_c)_{n',k'} \rbrace A_{n+n'+1,k+k'}  ) = \\
& =  \sum_{n,k,n',k'=0}^\infty P^{(2,2)}_{n',k',n,k} \sqrt{n+n'+1}\, \epsilon^{acb} \tr ( \lbrace (\zeta^\dagger_b)_{n,k} ,(\zeta^\dagger_c)_{n',k'} \rbrace A_{n+n'+1,k+k'}  ) = \\
& = - \sum_{n,k,n',k'=0}^\infty P^{(2,2)}_{n,k,n',k'} \sqrt{n+n'+1}\, \epsilon^{abc} \tr ( \lbrace (\zeta^\dagger_b)_{n,k} ,(\zeta^\dagger_c)_{n',k'} \rbrace A_{n+n'+1,k+k'} ) = 0 \, .
\end{aligned}
\eeq
In the first step, we used the symmetry of the anticommutator of fermions and we exchanged the labels $(n,k) \leftrightarrow (n',k')$ and $b \leftrightarrow c.$
In the last step, we used the symmetry properties \eqref{eq:useful_identities_P-Qsupercharge} and the antisymmetry of the Levi-Civita symbol.
Combining all the terms, we then find
\beq
 \lbrace Q_{4-a}, \mathcal{Q} \rbrace_D = 0 \, . 
\eeq

%%%%%%%%%%%%%%%%%%
\section{Conventions and details of the spherical expansion}
\label{app:sphere_red}

In this Appendix we collect the essential conventions used to apply the spherical expansion procedure in Section \ref{sec:ham_sphere}.
We omit several details which are not necessary for the understanding of the results derived in the present work.
The interested reader can find several other details in the following references:
\begin{itemize}
\item Appendix A of \cite{Baiguera:2020jgy}: conventions on the expansion of all the fields into spherical harmonics.
\item Appendix B of \cite{Baiguera:2020jgy}: technical details about the treatment of fermionic modes, computation of the Cartan charges of the $\mathcal{N}=4$ SYM action, weights associated to the fields.
\item Appendix A of \cite{Baiguera:2020mgk}: crossing relations between the Clebsch-Gordan coefficients on the three-sphere, general methods to perform the summations entering the computation of the interacting Hamiltonian.
\end{itemize}

\subsection{Definition of the Clebsch-Gordan coefficients}
\label{app-definition_Clebsch}

We start from the list of Clebsch-Gordan coefficients on $S^3.$
Their explicit expressions read
\beq
\mathcal{C}^{J_1 M_1}_{J_2 M_2; JM} = 
\sqrt{\frac{(2J+1)(2J_2+1)}{2J_1+1}}
C^{J_1 m_1}_{J_2 m_2; J m} C^{J_1 \tilde{m}_1}_{J_2 \tilde{m}_2; J \tilde{m}} \, ,
\label{eq:app_definitionClebsch_C}
\eeq
\beq
\begin{aligned}
\mathcal{D}^{J_1 M_1}_{J_2 M_2 \rho_2; JM \rho} & = 
(-1)^{\frac{\rho_2+\rho}{2} +1}
\sqrt{3(2J_2+1)(2J_2+2 \rho_2^2 +1)(2J+1)(2J+2 \rho^2 +1)} \\
& \times C^{J_1,m_1}_{Q_2,m_2; Q,m} C^{J_1, \tilde{m}_1}_{\tilde{Q}_2, \tilde{m}_2 ; \tilde{Q},\tilde{m}}
\begin{Bmatrix}
Q_2 & \tilde{Q}_2 & 1 \\
Q & \tilde{Q} & 1 \\
J_1 & J_1 & 0  
\end{Bmatrix} \, ,
\label{eq:app_definitionClebsch_D}
 \end{aligned}
\eeq
\beq
\begin{aligned}
 \mathcal{E}_{J_1 M_1 \rho_1; J_2 M_2 \rho_2; JM \rho} & =  \sqrt{6(2J_1+1)(2J_1 + 2 \rho_1^2 +1)(2J_2+1)(2J_2+2 \rho_2^2 +1)(2J+1)(2J+2 \rho^2 +1)} \\
& \times (-1)^{-\frac{\rho_1+\rho_2+\rho+1}{2}} 
\begin{Bmatrix}
Q_1 & \tilde{Q}_1 & 1 \\
Q_2 & \tilde{Q}_2 & 1 \\
Q & \tilde{Q} & 1 
\end{Bmatrix} 
\begin{pmatrix}
Q_1 & Q_2 & Q \\
m_1 & m_2 & m
\end{pmatrix}
\begin{pmatrix}
\tilde{Q}_1 & \tilde{Q}_2 & \tilde{Q} \\
\tilde{m}_1 & \tilde{m}_2 & \tilde{m}
\end{pmatrix} \, ,
\label{eq:app_definitionClebsch_E}
\end{aligned}
\eeq
\beq
\begin{aligned}
\mathcal{F}^{J_1 M_1 \kappa_1}_{J_2 M_2 \kappa_2; JM} = & (-1)^{\tilde{U}_1+U_2+J+\frac{1}{2}} \sqrt{(2J+1)(2J_2+1)(2J_2+2)}  \\
& \times C^{U_1, m_1}_{U_2, m_2; J,m} C^{\tilde{U}_1,\tilde{m}_1}_{\tilde{U}_2,\tilde{m}_2; J,\tilde{m}}
\begin{Bmatrix}
U_1 & \tilde{U}_1 & \frac{1}{2} \\
\tilde{U}_2 & U_2 & J
\end{Bmatrix} \, ,
\label{eq:app_definitionClebsch_F}
\end{aligned}
\eeq
\beq
\begin{aligned}
\mathcal{G}^{J_1 M_1 \kappa_1}_{J_2 M_2 \kappa_2; JM \rho} = & (-1)^{\frac{\rho}{2}} \sqrt{6(2J_2+1)(2J_2+2)(2J+1)(2J+2 \rho^2 +1)} \\
& \times C^{U_1,m_1}_{U_2,m_2; Q,m} C^{\tilde{U}_1, \tilde{m}_1}_{\tilde{U}_2, \tilde{m}_2 ; \tilde{Q},\tilde{m}}
\begin{Bmatrix}
U_1 & \tilde{U}_1 & \frac{1}{2} \\
U_2 & \tilde{U}_2 & \frac{1}{2} \\
Q & \tilde{Q} & 1 
\end{Bmatrix} \, ,
\label{eq:app_definitionClebsch_G}
\end{aligned}
\eeq
where we defined the quantities
\beq
U \equiv J + \frac{\kappa+1}{4} \spa
\tilde{U} \equiv J + \frac{1-\kappa}{4} \spa
Q \equiv J+ \frac{\rho(\rho+1)}{2} \spa
\tilde{Q} \equiv J + \frac{\rho(\rho-1)}{2} \, ,
\label{eq:app_labels_harmonics}
\eeq
with the labels running over $\kappa= \pm 1/2$ and $\rho \in \lbrace -1, 0, 1\rbrace .$

\subsection{Interacting Hamiltonian of $\mathcal{N}=4$ SYM upon spherical expansion}

We list the full interacting Hamiltonian of $\mathcal{N}=4$ SYM derived upon performing spherical expansion, before restricting to any near-BPS limit.
The computation was presented in reference \cite{Ishiki:2006rt}, but here we report the result using a notation consistent with our conventions for the fields.
In order to obtain a more compact expression, we use the definition
\beq
(Z_a)_{JM}  \equiv \begin{pmatrix} (\Phi_1)_{JM} \\
 (-1)^{m-\tilde{m}} (\Phi^\dagger_2)_{J,-M} \\
  (-1)^{m-\tilde{m}} (\Phi^\dagger_3)_{J,-M} 
  \end{pmatrix} \, ,
  \label{eq:mapping_Z_scalars}
\eeq
\beq
(\Psi_A)_{J,M,\kappa=1} \equiv   (\psi^{\dagger}_A)_{J,-M,\kappa=1} \, , \qquad
(\Psi_A)_{J,M,\kappa=-1} \equiv (\psi_A)_{J,M, \kappa=-1} \, .
\label{eq:mapping_Psi_fermions}
\eeq 
The origin for this choice resides in the comparison between the conventions adopted in \cite{Ishiki:2006rt} and in Appendix B of \cite{Baiguera:2020jgy}.
In this way, the interacting Hamiltonian reads:
\beq
\begin{aligned}
H_{\rm int}   = \sum_{J_i, M_i, \kappa_i, \rho_i} & \tr  \left\lbrace i g {\cal C}^{J_2M_2}_{J_1M_1;JM} \, \chi_{JM} \left([(Z_a^\dagger)_{J_2M_2},(\Pi^{\dagger}_a)_{J_1M_1}] + [(Z^a)_{J_1M_1},(\Pi^{a})_{J_2M_2}]\right)  \right. \\
& \left. - 4g\sqrt{J_1(J_1+1)} {\cal D}^{J_2M_2}_{J_1M_1 0; JM\rho} \, A_{(\rho)}^{JM}
[(Z^a)_{J_1M_1},(Z_a^\dagger)_{J_2M_2}]  \right. \\
& \left. +g \mathcal{F}^{J_1 M_1 \kappa_1}_{J_2 M_2 \kappa_2; JM}  \, \chi_{JM} 
\lbrace (\Psi_A^{\dagger})_{J_1 M_1 \kappa_1} , (\Psi^A)_{J_2 M_2 \kappa_2} \rbrace \right. \\
& \left. + g \mathcal{G}^{J_1 M_1 \kappa_1}_{J_2 M_2 \kappa_2; JM \rho} \, A_{(\rho)}^{JM}
\lbrace (\Psi^{\dagger}_A)_{J_1 M_1 \kappa_1} , (\Psi^A)_{J_2 M_2 \kappa_2} \rbrace   
  \right. \\
  & \left. + \frac{g^2}{2} {\cal C}^{J_2 M_2}_{J_1 M_1;J M}{\cal C}^{J_3 M_3}_{J_4 M_4;J M}[(Z^a)_{J_1 M_1},(Z^{\dagger}_a)_{J_2 M_2}][(Z^b)_{J_3 M_3},(Z_b^{\dagger})_{J_4 M_4}] \right. \\
 & \left.   - \sqrt{2} ig (-1)^{-m_1+\tilde{m}_1+\frac{\kappa_1}{2}} \mathcal{F}^{J_1,-M_1,\kappa_1}_{J_2 M_2 \kappa_2; J M} 
 \Psi^4_{J_2 M_2 \kappa_2} [(Z_a)^{JM} , (\Psi^a)_{J_1 M_1 \kappa_1}]  \right. \\
 & \left.   + \sqrt{2}ig (-1)^{-m_1+\tilde{m}_1+\frac{\kappa_1}{2}} \mathcal{F}^{J_1,-M_1,\kappa_1}_{J_2 M_2 \kappa_2; J M} 
\, \epsilon^{abc} (\Psi_a)_{J_1 M_1 \kappa_1} [(Z_b^{\dagger})^{JM} , (\Psi_c)_{J_2 M_2 \kappa_2}]  \right. \\
  & \left.   + \sqrt{2}ig (-1)^{m_2-\tilde{m}_2+\frac{\kappa_2}{2}} \mathcal{F}^{J_1 M_1 \kappa_1}_{J_2, -M_2, \kappa_2; J M} 
 (\Psi^{\dagger}_4)_{J_2 M_2 \kappa_2} [(Z^{\dagger}_a)^{JM} , (\Psi^{\dagger}_a)_{J_1 M_1 \kappa_1}]  \right. \\
 & \left.   - \sqrt{2}ig (-1)^{m_2-\tilde{m}_2+\frac{\kappa_2}{2}} \mathcal{F}^{J_1 M_1 \kappa_1}_{J_2, -M_2, \kappa_2; J M} 
\, \epsilon^{abc} (\Psi^{\dagger}_a)_{J_1 M_1 \kappa_1} [(Z_b)^{JM} , (\Psi^{\dagger}_c)_{J_2 M_2 \kappa_2}]  \right. \\
& \left. + i g {\cal D}^{JM}_{J_1M_1\rho_1;J_2M_2\rho_2} \, \chi_{JM} [\Pi^{J_1M_1}_{(\rho_1)},A^{J_2M_2}_{(\rho_2)}] \right. \\
& \left. + g^2 \mathcal{C}^{JM}_{J_2 M_2; J_4, -M_4} \mathcal{D}_{JM; J_1 M_1 \rho_1; J_3 M_3 \rho_3} [A_{(\rho_1)}^{J_1 M_1} , (Z^a)_{J_2 M_2}] [A_{(\rho_3)}^{J_3 M_3} , (Z^{\dagger}_a)_{J_4 M_4}]   \right. \\
& \left. + 2 i g \rho_1(J_1 + 1) {\cal E}_{J_1M_1\rho_1;J_2M_2\rho_2;J_3M_3\rho_3}A^{J_1M_1}_{(\rho_1)}[A^{J_2M_2}_{(\rho_2)},A^{J_3M_3}_{(\rho_3)}] \right. \\
& \left. - \frac{g^2}{2} {\cal D}^{JM}_{J_1M_1\rho_1; J_3M_3\rho_3} {\cal D}_{JM;J_2M_2\rho_2; J_4 M_4 \rho_4}[A^{J_1M_1}_{(\rho_1)},A^{J_2M_2}_{(\rho_2)}][A^{J_3M_3}_{(\rho_3)},A^{J_4M_4}_{(\rho_4)}] \right. \\
& \left. -2g \sqrt{J_1(J_1+1)} \mathcal{D}_{J_2 M_2; J_1 M_1 0; J M \rho} \, \chi_{J_1 M_1} [\chi_{J_2 M_2} , A_{(\rho)}^{JM}] \right. \\
& \left. + \frac{g^2}{2} \mathcal{C}^{JM}_{J_1 M_1; J_3 M_3} \mathcal{D}_{JM; J_2 M_2 \rho_2; J_4 M_4 \rho_4} [\chi_{J_1 M_1} , A_{(\rho_2)}^{J_2 M_2}] [\chi_{J_3 M_3} , A_{(\rho_4)}^{J_4 M_4}]  \right. \\
& \left. + g^2 \mathcal{C}^{JM}_{J_1 M_1; J_2 M_2} \mathcal{C}_{JM; J_3 M_3; J_4 M_4} [\chi_{J_1 M_1} , (Z^a)_{J_2M_2}] [\chi_{J_3 M_3} , (Z_a^{\dagger})_{J_4 M_4}]
 \right\rbrace \, .
\end{aligned}
\label{eq:app_full_interacting_N=4SYM_Hamiltonian}
\eeq
We add few comments on the notation:
\begin{itemize}
\item The overall summation over the contracted indices involves momenta $(J,M),$ labels for fermions $(\kappa)$ and gauge fields $(\rho),$ and indices $a,A$ of the various fields under $\SU{(4)}$ R-symmetry. 
\item The fermions summed in the Yukawa term run over $a \in \lbrace 1,2,3  \rbrace$ and the corresponding Levi-Civita symbol is defined in such a way that $\epsilon^{123}=1.$
\item $\Pi_a$ are the canonical momenta associated to the scalar fields $\Phi_a,$ while $\Pi_{(\rho)}$ is the symplectic partner of the gauge field $A_{(\rho)}.$
\end{itemize}

\addcontentsline{toc}{section}{References}

\bibliography{newbib}
\bibliographystyle{newutphys}

\end{document}